%% file: Efficient_Trotterizations.tex
\documentclass[12pt]{article}
\usepackage[english]{babel}
\usepackage{fullpage}
\usepackage{amsfonts}
\usepackage{wrapfig}
\usepackage{amsmath}
\usepackage{booktabs}
\usepackage{multirow}
\usepackage{placeins}
\usepackage{pgf}
\usepackage{nicefrac}
\usepackage{authblk}
\usepackage[binary-units=true]{siunitx}
\usepackage{amssymb}
\usepackage{array}
\usepackage{graphicx}
\usepackage{dsfont}
\newcolumntype{C}[1]{>{\centering\arraybackslash}m{#1}}
\newcolumntype{R}[1]{>{\raggedleft\arraybackslash}m{#1}}
\usepackage{caption}
\usepackage[backend=bibtex8,style=phys,sorting=none,citestyle=numeric-comp,eprint=true,isbn=true]{biblatex}
\usepackage{hyperref}
\usepackage{cleveref}
\hypersetup{colorlinks=true}

\definecolor{slateblue}{HTML}{31207F}
\definecolor{darkviolet}{HTML}{7B207F}
\definecolor{crimson}{HTML}{7F1132}

\hypersetup{
    colorlinks=true,
    linkcolor=slateblue,
    citecolor=darkviolet,
    urlcolor=crimson
}

\DeclareMathOperator{\ord}{\mathcal{O}}
\newcommand{\order}[1]{\ensuremath{\ord\left(#1\right)}}

\author{Marko Maležič}
\author{Johann Ostmeyer}
\affil{Helmholtz-Institut f\"ur Strahlen- und Kernphysik, University of
	Bonn, Nussallee 14-16, 53115 Bonn, Germany}
\title{Efficient Trotter-Suzuki Schemes\\ for Long-time Quantum Dynamics}

\begin{document}
\maketitle

\begin{abstract}
	Accurately simulating long-time dynamics of many-body systems is a challenge in both classical and quantum computing due to the accumulation of Trotter errors.
	While low-order Trotter-Suzuki decompositions are straightforward to implement, their rapidly growing error limits access to long-time observables.
  We present a framework for constructing efficient high-order Trotter-Suzuki schemes by identifying their structure and directly optimizing their parameters over a high-dimensional space.
	This method enables the discovery of new schemes with significantly improved efficiency compared to traditional constructions, such as those by Suzuki and Yoshida.
	Based on the theoretical efficiency and practical performance, we recommend two novel highly efficient schemes at $4^{\textrm{th}}$ and $6^{\textrm{th}}$ order.
  We also demonstrate the effectiveness of these decompositions on the Heisenberg model and the quantum harmonic oscillator, and find that for a fixed final time they perform better across the computational cost.
  Even when using large time steps, they surpass established low-order schemes like the Leapfrog.
  Finally, we investigate the in-practice performance of different Trotter schemes and find the decompositions with more uniform coefficients tend to feature improved error accumulation over long times.
  We have included this observation into our choice of recommended schemes.
\end{abstract}

\newpage

\input{intro}

\input{framework}

\input{novel_schemes}

\input{uniformity}

\input{conclusion}

\section*{Code and Data}

All the code needed to reproduce the results in this work can be found on Github or Zenodo~\cite{markomalezic_2026_18347430}.
The derivations for the recursive formulae were done with the help of Wolfram Mathematica~\cite{Mathematica}, with a dependency on routines provided by FeynCalc~\cite{MERTIG1991345, Shtabovenko2025}.
We also used Mathematica to calculate and visualize the polynomial manifolds in \Cref{sec:optim}.
The code for evaluating Trotter-Suzuki schemes and the minimizer for finding optimal scheme parameters is written in \texttt{C++}, with a dependency on the linear algebra library Eigen~\cite{eigenweb}.
The data for the schemes found by the minimizer can be accessed on Github/Zenodo.
Time evolution routines for the quantum harmonic oscillator (see \Cref{sec:novel_schemes:Harmonic}) are written in Python, and the dynamics of the Heisenberg model (see \Cref{sec:novel_schemes:Heisenberg}) are again written is \texttt{C++}, with a dependency on Eigen.
The data for the time evolution can be reproduced with the given code, but it is slightly too large for publication on our repository.
Nonetheless, the data will be gladly provided upon request.

\section*{Acknowledgements}
We thank Anthony Kennedy, Paul Ludwig, Emanuele Mendicelli, Benjamin Søgaard and Lorenzo Verzichelli for insightful discussions.
The authors gratefully acknowledge the access to the Marvin cluster of the University of Bonn.
This work was funded by the Deutsche Forschungsgemeinschaft (DFG, German Research Foundation) as part of the CRC 1639 NuMeriQS – Project number 511713970.

%\clearpage
\FloatBarrier
\printbibliography

\appendix

\input{framework_details}

\input{error_accumulation}

\end{document}

%% file: intro.tex
% !TEX root = Efficient_Trotterizations

\section{Introduction}\label{sec:intro}

Understanding the time evolution of classical and quantum many-body systems has remained a difficult task in many fields of physics.
Rarely are these systems integrable and provide an exact solution to their dynamics.
We are thus forced to make certain compromises in order to extract physical properties of complex systems.
We list some systems of interest in contemporary physics in the following subsection.
The compromises we make usually rely on numerical simulations and the trade-off made is commonly the loss of accuracy.
In this paper, we investigate the compromise made when approximating an exponential of sums of non-commuting generators~$A_{i}$, more commonly known as Trotter-Suzuki product formulas or splitting methods,

\begin{equation}
  \exp \left[ \sum\nolimits_{i} A_{i} h + \mathcal{O} \left( h^{2} \right) \right] = \prod_{i} e^{A_{i} h}. \label{eq:trotter}
\end{equation}

\noindent
Here we have written the simplest product formula for an arbitrary number of operators~$A_{i}$; this incurs a Trotter error of~$\mathcal{O} \left( h^{2} \right)$ in the parameter~$h$.
We study how more involved product formulae improve the scaling of the Trotter error, and how they can be used to improve simulations of time evolution.

\subsection{Time evolution}

Relations like the one in~\eqref{eq:trotter} are used widely in time evolution, and possibly known better in simulations of quantum systems.
In quantum mechanics, the time evolution of a state~$| \psi (t) \rangle$ is governed by the Schrödinger equation,

\begin{equation}
  \frac{d}{d t} | \psi (t) \rangle = - i H | \psi (t) \rangle \; \longrightarrow \; | \psi (t) \rangle = e^{-it H} | \psi(0) \rangle,
\end{equation}

\noindent
where we set $\hbar = 1$, and identify the time evolution operator $U (-it) = e^{-it H}$ by exponentiation of the Hamiltonian~$H$.
An analytic solution for the time evolution operator is not always available, in which case one has to resort to numerical diagonalization of the Hamiltonian~$H$.
However, this approach fails when the Hilbert space of the system grows exponentially.
If the Hamiltonian can be expressed as a sum of local terms $H = \sum_{i} A_{i}$, then it is possible to get over the scalability problem by employing Trotter-Suzuki decompositions~\eqref{eq:trotter} on the time evolution operator~$U$.
Many physically important local systems have yet to be fully explored due to their complexity, e.g.\ lattice gauge theories in the Hamiltonian formulation~\cite{Jakobs:2025rvz, Kane:2025ybw, Fontana:2024rux, Jakobs:2025zcv} or condensed matter models both in equilibrium as well as far from it~\cite{RevModPhys.68.13, RevModPhys.86.779, RevModPhys.83.863, Eisert_2015}.
Recently much work has been done to prepare such simulations to be executed on quantum hardware, where Trotterizations are used for time evolution with unitarity preservation.
It is our hope that with efficient Trotter-Suzuki decompositions, we can improve on the accuracy of such simulations.
Though a significant part of the paper focuses on unitary time evolution, we note another method, which does not rely on locally expressed Hamiltonians.
It is possible to expand an exponential like the time evolution operator into a Taylor or Chebyshev series~\cite{Ostmeyer:2023xju}.
Both can approximate time evolution very accurately, but they are non-unitary and therefore cannot be used in simulations that require exact unitarity (e.g.\ on quantum computers).

Because Trotter-Suzuki schemes represent general operator exponentials they are also useful in simulations of certain classical systems, like those with the following Hamiltonian form,

\begin{equation}
  H (\tilde{p}, \tilde{q}) = T(\tilde{p}) + V(\tilde{q}), \quad T(\tilde{p}) = \frac{\tilde{p}^{2}}{2 m},
\end{equation}

\noindent
written as a sum of the kinetic part~$T(\tilde{p})$ and the potential part~$V(\tilde{q})$, which depend on the generalized momenta~$\tilde{p}$ and coordinates~$\tilde{q}$ respectively.
In the Liouville formulation, the solution for a phase-space distribution~$f(\tilde{p}, \tilde{q}, t)$ is found in a similar way,

\begin{equation}
  \frac{\partial f}{\partial t} = \left\{ H, f \right\} = (L_{T} + L_{V}) f \; \longrightarrow \; f (t) = e^{(L_{T} + L_{V}) t} f(0),
\end{equation}

\noindent
where we defined $L_{T} = \left\{ T, \cdot \right\}$ and $L_{V} = \left\{ V, \cdot \right\}$ with Poisson brackets.

In this formulation, Trotter-Suzuki decompositions are applicable for the numerical solution of the equations of motion (EOM)

\begin{equation}
	\frac{d \tilde{q}}{d t} = \frac{\partial H}{\partial \tilde{p}}, \quad \frac{d \tilde{p}}{d t} = - \frac{\partial H}{\partial \tilde{q}}.
\end{equation}

\noindent
These schemes are often called Runge-Kutta-Nyström methods or symplectic integrators, and much work has been done by Omelyan \textit{et al.}~\cite{OMELYAN2003272} to construct such schemes.
The optimization of symplectic integrators is subtly different from that of general Trotterizations because the nested commutator $[L_V, [L_V, [L_V, L_T]]] = 0$ vanishes.
The molecular dynamics EOM may show up in many classical fields of physics, notably however they are used in simulations of large molecules in chemistry~\cite{doi:10.1021/acs.jcim.4c00823} and biophysics~\cite{Hollingsworth2018}, or many small molecules in soft matter physics~\cite{WONGEKKABUT20162529}.
They are also useful in studies of astrophysics~\cite{10.1093/mnras/stv1439, Beust2003}, but surprisingly they also show up in stochastic algorithms like the Hybrid Monte Carlo~\cite{Duane:1987de,Ostmeyer:2025agg}.
Although, this paper does not focus specifically on improving these integrators, we believe that the framework we build could be extended to explore this avenue as well.

\subsection{Trotter-Suzuki decompositions}

Trotterizations or Trotter-Suzuki decomposition schemes are approximations of operator exponentials, such as the one in Eq.~\eqref{eq:trotter}, and we can denote them as $S_{n} (h)$.
Every scheme approximates the true exponent $S_{n} (h) = U(h) + \mathcal{O} (h^{n+1})$ up to an order~$n$ in its step size~$h$.
Due to the Lie-Trotter formula~\cite{lie1888theorie,trotter_original} and subsequent work by Suzuki~\cite{Suzuki:1976be}, it is known that any such scheme $S_{n} (h)$ converges to the correct exponential operator in the limit $h \rightarrow 0$.
Suzuki also proposed a method to elevate any scheme~$S_{n}$ to a higher order $n + 2$, see Ref.~\cite{Hatano_2005}.
Similarly, Yoshida~\cite{YOSHIDA1990262} derived a more involved approach to construct higher order decompositions from symmetric schemes of lower orders.
Both of these methods however exhibit a fundamental flaw---the restriction on known low order schemes, which manifests itself in a bigger Trotter error, although its scaling is improved.
Without the need to rely on lower orders, it is possible to construct schemes from scratch, which is the method pioneered by Omelyan \textit{et al.}~\cite{OMELYAN2003272}.
This is also how we approach the problem in this paper.
Although conceptually more complicated, it has the potential to greatly reduce the Trotter error within a given order~$n$.
Some highly efficient schemes were derived by Omelyan \textit{et al.}~\cite{Omelyan_2002,OMELYAN2003272} at orders $n = 2, 4$ as well as orders $n = 4, 6$ by Blanes and Moan~\cite{BLANES2002313}.
In this paper, we build a framework for scheme construction based on this method and propose improved schemes for orders $n = 4, 6$.
The framework could be extended to higher orders $n \geq 8$ and used to improve on the currently best schemes at these orders by Morales \textit{et al.}~\cite{morales2022greatly}, which are based on Yoshida's method.

The theoretical efficiency we use for the optimization of Trotter decompositions is based on an approximation of the single time step error without prior knowledge of the operators.
Some work has already been done to tighten the upper bounds of Trotter errors for general product formulae including information about the operators explicitly~\cite{PhysRevX.11.011020, schubert2023trotter}.
However, this estimate seems to be quite model dependent and the bounds remain a rather loose approximation.
This is partially due to completely neglecting the interference of multiple Trotter steps.
Besides their error of a single time step, Trotterizations exhibit another important aspect, which is the practical error accumulation.
The latter half of this paper focuses specifically on this property of Trotter-Suzuki schemes.
The recent Refs.~\cite{chen2024trottererrortimescaling, chen2024errorinterferencequantumsimulation} provide some intuition behind the time scaling and how the bounds can be improved.
Based on the findings in~\cite{chen2024trottererrortimescaling} and some experimental observations we propose an addition to the framework.
The idea is to minimize the part of the Trotter error that interferes constructively over multiple steps.
To some extent, this can be achieved by minimizing the variance of the Trotter coefficients.
Unfortunately, this ansatz only gives minimal improvement in the error accumulation.

%% file: framework.tex
% !TEX root = Efficient_Trotterizations

\section{The framework of optimized Trotterizations}\label{sec:framework}
Trotterizations provide a systematic way to approximate exponentials of operators. Consider the minimal example

\begin{equation}
	e^{A h} e^{B h} = e^{(A + B) h + \mathcal{O} (h^{2})}
\end{equation}

\noindent
in the case of two non-commuting operators $A$ and $B$, where the approximation leads to a local error of order~$\mathcal{O} (h^{2})$ in the parameter~$h$.
We can study such approximations with an arbitrary number of operators $A_{k}$ up to a desired order~$\mathcal{O} (h^{n+1})$, where we denote the order as~$n$.
Though these approximations are interesting mathematically by themselves, they are commonly found in time evolution of quantum systems, which is why we will adopt naming conventions motivated by physics throughout this work.
Such systems are described by the Hamiltonian $H$ and their evolution in (imaginary) time $t$ is given by the time evolution operator $U(t) = \exp(t H)$.
In many complex systems the time evolution operator cannot be evaluated analytically, which is why Trotterizations are used to approximate it:  $U(t) \approx \left[ S_{n} (h) \right]^{t/h}$, where $S_{n}$ represents a decomposition of order $n$ split into $N_{t} = t/h$ time steps of length $h$.
As the system evolves in time according to the Trotterization $S_{n}$, it accumulates the local errors of order $\mathcal{O}(h^{n+1})$ at each step up to a global error which scales as $\mathcal{O}(h^{n+1} N_{t}) = \mathcal{O} (t h^{n})$.

\begin{figure}[h!]
	\begin{center}
		\includegraphics[width=0.9\textwidth]{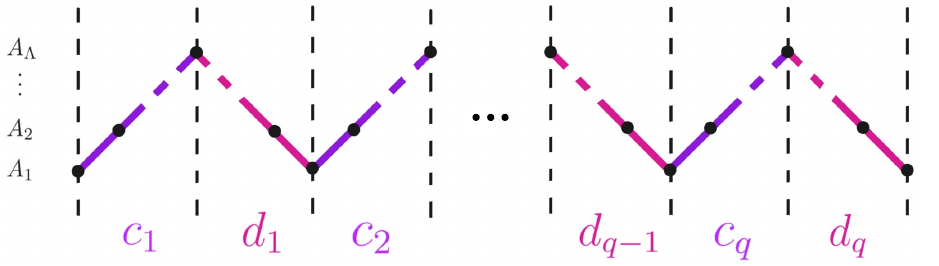}
	\end{center}
	\caption{Representation of a Trotter-Suzuki scheme with an arbitrary number of stages $\Lambda$.
		There are $q$ cycles, each consisting of two ramps.
		Ramps forward are indicated by the purple line, while pink lines represent the backward ramps.
		Read from left to right, while multiplying exponents of operators $A_{k}$ with appropriate parameters $c_{i}$ or $d_{i}$, one obtains the decomposition from Eq.~\eqref{eq:general_scheme}.
		Adapted from~\cite{Ostmeyer:2022}.}\label{fig:ramp}
\end{figure}

In the following we use a notation similar to the one introduced by Omelyan \textit{et al.}~\cite{Omelyan_2002}, and review some theoretical aspects of the framework also applied in~\cite{Ostmeyer:2022}.
For a Hamiltonian consisting of multiple operators $A_{k}$: $H = \sum_{k=1}^{\Lambda} A_{k}$, the time evolution operator can be split into many \textit{stages}: $e^{c_{i} h A_{k}}$ and $e^{d_{i} h A_{k}}$, over \textit{sub-steps} $c_{i} h$ and $d_{i} h$, where parameters $c_{i}$ and $d_{i}$ define the specific decomposition $S_{n}$.
All $\Lambda$ stages create either a ramp forward $\prod_{k=1}^{\Lambda} e^{c_{i} h A_{k}}$ or a ramp backward $\prod_{k=\Lambda}^{1} e^{d_{i} h A_{k}}$ (mind the order of the $k$ indices).
Completing a forward and backward ramp defines a \textit{cycle} and a single step is composed of $q$ cycles, which uniquely defines a scheme $S_{n}$, and its order $n$.
\Cref{fig:ramp} provides a visualization of these concepts, which produce a general formula for a decomposition scheme:

\begin{equation}
	S_{n} (h) = \left( \prod_{k = 1}^{\Lambda} e^{c_{1} h A_{k}} \right) \left( \prod_{k = \Lambda}^{1} e^{d_{1} h A_{k}} \right) \cdots \left( \prod_{k = 1}^{\Lambda} e^{c_{q} h A_{k}} \right) \left( \prod_{k = \Lambda}^{1} e^{d_{q} h A_{k}} \right), \label{eq:general_scheme}
\end{equation}

\noindent
where we can see that the number of parameters $c_{i}$ and $d_{i}$ of a decomposition is equal to $2 q$.
This is the so-called ramp-based approach, which is useful in practice.

The goal in deriving and optimizing Trotter-Suzuki schemes is finding the optimal parameters $c_{i}$ and $d_{i}$ for a given number of cycles $q$ and desired order $n$.
Doing so for an arbitrary number of stages has recently been made relatively simple, since proofs in Refs.~\cite{McLachlan:1995otni, Ostmeyer:2022} provide a straightforward transformation between parameters in 2-stage and arbitrary stage decompositions.
This allows us to derive schemes in the simpler case of 2 operators $A$ and $B$, which has the following form:

\begin{equation}
	e^{h (A + B) + \mathcal{O} (h^{n+1})} = e^{\mathsf{a}_{1} h A} e^{\mathsf{b}_{1} h B} e^{\mathsf{a}_{2} h A} \cdots e^{\mathsf{b}_{q} h B} e^{\mathsf{a}_{q+1} h A},\label{eq:2stage}
\end{equation}

\noindent
which is written down in the stage-based approach.
This approach is convenient for derivations, but impractical in reality.
We can apply the found parameters $\mathsf{a}_{i}$ and $\mathsf{b}_{i}$ to a complex Hamiltonian with many non-commuting operators, according to the transformation 

\begin{equation}
  \begin{aligned}
	  c_{1} &= \mathsf{a}_{1},           \hspace{0.1\linewidth} &d_{1} = \mathsf{b}_{1} - c_{1},                   \\
	  c_{2} &= \mathsf{a}_{2} - d_{1},   \hspace{0.1\linewidth} &d_{2} = \mathsf{b}_{2} - c_{2},                   \\
	  &\hspace{0.225cm}\vdots              \hspace{0.1\linewidth} &\vdots\hspace{1.525cm}                              \\
	  c_{i} &= \mathsf{a}_{i} - d_{i-1}, \hspace{0.1\linewidth} &d_{i} = \mathsf{b}_{i} - c_{i}, \hspace{0.075cm}  \\
	  &\hspace{0.225cm}\vdots              \hspace{0.1\linewidth} &\vdots\hspace{1.525cm}                              \\
	  c_{q} &= \mathsf{a}_{q} - d_{q-1}, \hspace{0.1\linewidth} &d_{q} = \mathsf{b}_{q} - c_{q}.
  \end{aligned}
\end{equation}

\noindent
Using these parameters $c_{i}$ and $d_{i}$, Eq.~\eqref{eq:2stage} can be rewritten to obtain the product formula for a general number of stages given in equation~\eqref{eq:general_scheme}.
Besides this simplification, we shall focus solely on symmetric decomposition schemes, which are defined by symmetrizing the parameters in Eq.~\eqref{eq:2stage}, i.e. $\mathsf{a}_{1} = \mathsf{a}_{q+1}$, $\mathsf{b}_{1} = \mathsf{b}_{q}$, $\ldots$
The reason behind this is, that any non-symmetric scheme can be elevated to an even order $n$ without any drawbacks, by reverting the sequence of parameters $\mathsf{a}_{i}$ and $\mathsf{b}_{i}$ in every second step.
The induced symmetric scheme with even order is strictly more efficient than the initial one, therefore we specifically focus on even orders $n$.

A few methods have been proposed to obtain higher order schemes, which differ fundamentally in their approaches.
Both Suzuki~\cite{Suzuki:1976be,Hatano_2005} and Yoshida~\cite{YOSHIDA1990262} derived methods, which use existing schemes of lower orders as building blocks to obtain a higher order scheme.
However, both fail to find the maximally efficient schemes at given cycles $q$ and desired order $n$.
In search of better efficiency Omelyan, Mryglod and Folk~\cite{OMELYAN2003272} as well as Blanes and Moan~\cite{BLANES2002313} constructed their schemes from scratch, which is significantly more complicated.
In return, the $4^{th}$ and $6^{th}$ order schemes they found had a better theoretical efficiency and performed better in practice than what had been known before.
Following in their footsteps, we construct a framework to derive Trotter-Suzuki schemes at given cycles $q$, which can be, in principle, expanded to any even order $n$.

\subsection{The error coefficients}

In order to derive a decomposition from scratch, we firstly write down its stage-based form.
Since we are working in the stage-based approach, it is important whether the number of cycles $q$ is even or odd.
We write down a scheme for an even number of cycles $q$, though the form is similar for odd $q$,

\begin{equation}
	e^{(A + B) h + O_{1} h + O_{3} h^{3} + O_{5} h^{5} + \cdots} = e^{a_{\nicefrac{q}{2}+1} h A} e^{b_{\nicefrac{q}{2}} h B} \cdots e^{b_{1} h B} e^{2 a_{1} h A} e^{b_{1} h B} \cdots e^{b_{\nicefrac{q}{2}} h B} e^{a_{\nicefrac{q}{2}+1} h A}, \label{eq:even_symm_scheme}
\end{equation}

\noindent
where we used a different notation to that of Eq.~\eqref{eq:2stage}, which we will denote in terms of scheme parameters $a_{i}$ and $b_{i}$ instead of $\mathsf{a}_{i}$, $\mathsf{b}_{i}$ in order to distinguish them.
In general the scheme parameters are complex valued $a_{i}, b_{i} \in \mathds{C}$, and for the duration of this section we consider them as such.
The scheme is constructed such that it is symmetric in parameters $a_{i}$, $b_{i}$, and their indexing starts in the middle and grows outward.
Finally, the first parameter in the scheme $a_{1}$ is multiplied by~2.
These changes make the derivation slightly more convenient.
We also note here that the number of parameters $a_{i}$ and $b_{i}$ is equal to $q + 1$.

The left-hand side of Eq.~\eqref{eq:even_symm_scheme} is a sum of operators and higher order commutators in $A$ and $B$, due to the Baker-Campbell-Hausdorff (BCH) formula applied to the right-hand side of the equation.
The BCH formula is the central piece of Trotter-Suzuki decompositions, and defines the rules we need to follow in our derivation.
The symmetric BCH formula, which is applicable to our approach is given by the following equation,

\begin{equation}
	e^{\nicefrac{A}{2}} e^{B} e^{\nicefrac{A}{2}} = \exp \left(A + B + \frac{1}{4} \tilde{\alpha} \left[ A, \left[ A, B \right] \right] + \frac{1}{2} \tilde{\beta} \left[ B, \left[ B, A \right] \right] + \cdots \right), \quad \tilde{\alpha} = - \frac{1}{6}, \; \tilde{\beta} = \frac{1}{6},\label{eq:BCH_formula}
\end{equation}

\noindent
where it can be noticed that all commutators of even order vanish, which is the underlying reason why symmetric schemes automatically elevate the order of a scheme.
We can now tackle the form of operators in the left-hand side of Eq.~\eqref{eq:even_symm_scheme}.
Firstly, we write the sum of $A$ and $B$ to recover our result -- the Hamiltonian $H = A + B$.
Furthermore, we can expect operators $A$ and $B$ to be multiplied by parameters $a_{i}$ and $b_{i}$, so we write down the operator~$O_{1}$,

\begin{equation}
	O_{1} = (\nu - 1) A + (\sigma - 1) B, \quad \nu = 2 \sum_{i} a_{i}, \; \sigma = 2 \sum_{i} b_{i}.\label{eq:O1}
\end{equation}

\noindent
where we introduced coefficients $\nu (a_{i})$ and $\sigma (b_{i})$.
A scheme recovers the Hamiltonian $H$ only if $O_{1} = 0$, or in other terms $\nu = 2 \sum_{i} a_{i} = 1$ and $\sigma = 2 \sum_{i} b_{i} = 1$ (or similarly in the ramp-based approach, $\sum_{i} c_{i} = \sum_{i} d_{i} = \frac{1}{2}$).
Constraining the parameters $a_{i}$ and $b_{i}$ in this way constructs a valid scheme for use in time evolution.

Finally, we can now discuss the higher order operators $O_{3}$, $O_{5}$ and so on, up to a desired order $n$.
From the given BCH formula~\eqref{eq:BCH_formula} we can see that $O_{3}$ should have the following form,

\begin{equation}
	O_{3} = \alpha C_{1} + \beta C_{2}, \quad C_{1} = [A, [A, B]], \; C_{2} = [B, [B, A]],\label{eq:O3}
\end{equation}

\noindent
where we again introduce new coefficients $\alpha (a_{i}, b_{i})$ and $\beta (a_{i}, b_{i})$, which depend on all scheme parameters.
If we wish to eliminate contributions of $\mathcal{O} (h^{3})$ in Eq.~\eqref{eq:even_symm_scheme}, and construct an order $n = 4$ scheme, we need to set $O_{3} = 0$.
This is satisfied by the constraints: $\alpha (a_{i}, b_{i}) = \beta (a_{i}, b_{i}) = 0$.
We also renamed the commutators to $C_{1}$ and $C_{2}$, however this choice of commutators is not the only option.
While applying the BCH formula, one has to choose a basis of commutators, and there exist a few established bases like the Hall~\cite{Hall1934} or the Lyndon basis~\cite{Lyndon1958}.
Casas \textit{et al.}~\cite{Casas:2009,Arnal:2020xpt} later showed the Hall basis can be nicely transformed into a nested commutator basis, which is also maximally reduced using specific identities.
The choice of basis does not impact the scheme parameters themselves, but can impact the value of the theoretical efficiency, which may differ slightly between literature.
We followed the procedure by Casas \textit{et al.}\ closely and decided to choose a right-nested commutator basis, which is symmetric in the operators $A$ and $B$.
This can already be seen from equation~\eqref{eq:O3}, however let us now also write down the next order operator $O_{5}$ in the chosen basis,

\begin{equation}
  \begin{aligned}
    D_{1} &= [A, [A, [A, [A, B]]]], \quad D_{6} = [B, [B, [B, [B, A]]]],\\
    O_{5} = \sum\limits_{k=1}^{6} \gamma_{k} D_{k}, \quad D_{2} &= [A, [A, [B \hspace{-0.03cm}, [A, B]]]], \hspace{0.42cm} D_{5} = [B, [B, [A \hspace{0.03cm}, [B, A]]]],\\
    D_{3} &= [B \hspace{-0.03cm}, [A, [A, [A, B]]]], \hspace{0.4cm} D_{4} = [A \hspace{0.03cm}, [B, [B, [B, A]]]].
  \end{aligned}
\end{equation}

\noindent
Similarly to the previous orders, we define coefficients $\gamma_{k} (a_{i}, b_{i})$, which need to satisfy $\gamma_{k} (a_{i}, b_{i}) = 0$ if we wish to construct an order $n = 6$ scheme.
Higher orders are constructed in the same way, though increased orders bring increased complexity.
Bases of higher order operators also increase in size as well as become increasingly harder to define.
We therefore omit them here, but they can be found in our Mathematica notebook (\texttt{Efficient\_Trotterization.nb}) at~\cite{markomalezic_2026_18347430}.

\subsection{Recursive construction}\label{sec:recursive}

The goal in constructing higher order schemes in this approach is therefore to identify coefficients $\alpha, \beta, \gamma_{k}, \ldots$ and find the parameters $a_{i}$ and $b_{i}$, which constrain them to zero.
The way we approach this is by deriving recursive formulae, and using them to iteratively calculate the coefficients.
An iteration step in Eq.~\eqref{eq:even_symm_scheme} is an application of the symmetric BCH formula~\eqref{eq:BCH_formula} in the middle of the right-hand part of the equation.
First let us denote the result of the BCH formula at some point in the iteration as $\exp \left( \Phi^{(i_{A}, i_{B})} \right)$, where indices $i_{A}$ and $i_{B}$ count how many times the symmetric BCH has been used on stages $A$ or $B$.
The initial construction starts with $\Phi^{(0,0)} = 0$, and the symmetric BCH formula with stage $B$ is applied in the following way,

\begin{equation}
	\exp \left( b_{i_{B}} h B \right) \exp \left(  \Phi^{(i_{A}, i_{B}-1)} \right) \exp \left( b_{i_{B}} h B \right) = \exp \left(  \Phi^{(i_{A}, i_{B})} \right).\label{eq:BCH_B}
\end{equation}

\noindent
Following from Eq.~\eqref{eq:BCH_formula}, the expression $\Phi^{(i_{A}, i_{B})}$ must then have the following form,

\begin{equation}
	\Phi^{(i_{A}, i_{B})} = \left ( \nu^{(i_{A})} A + \sigma^{(i_{B})} B \right ) h + \left ( \alpha^{(i_{A}, i_{B})} C_{1} + \beta^{(i_{A}, i_{B})} C_{2} \right ) h^{3} + \cdots,
\end{equation}

\noindent
where $\nu^{(i_{A})}$ and $\sigma^{(i_{B})}$ sum up the parameters $a_{i}$, $b_{i}$ up to the iteration $(i_{A}, i_{B})$:

\begin{equation}
	\nu^{(i_{A})} = 2 \sum_{i=1}^{i_{A}} a_{i}, \quad \sigma^{(i_{B})} = 2 \sum_{i=1}^{i_{B}} b_{i}.
\end{equation}

We now show how to derive the recursive formulae for coefficients $\alpha$ and $\beta$ if the number of cycles $q$ is even, however the case of odd $q$ follows the same pattern.
Because our schemes follow an alternating application of the BCH formula on stages $A$ and $B$, indices $i_A$ and $i_B$ will always be either the same or separated by 1.
We therefore rename $i_{A}$ to $i$ and let $i_{B}$ be determined by it.
Applying the symmetric BCH relation~\eqref{eq:BCH_formula} for stage $B$ like in Eq.~\eqref{eq:BCH_B}, and collecting all terms in $h^{3}$, we can write down the following recursive formula:

\begin{align}
	\alpha_{B}^{(i, i)} C_{1} + \beta_{B}^{(i, i)} C_{2} &= \alpha_{A}^{(i, i-1)} C_{1} + \beta_{A}^{(i, i-1)} C_{2} + \tilde{\alpha} \left[ b_{i} B, \left[ b_{i} B, \nu^{(i)} A + \sigma^{(i-1)} B \right] \right] + \nonumber \\
	&\hspace{0.5cm} + \tilde{\beta} \left[\nu^{(i)} A + \sigma^{(i-1)} B, \left[ \nu^{(i)} A + \sigma^{(i-1)} B, b_{i} B \right] \right] = \nonumber \\
	&= \left[ \alpha_{A}^{(i, i-1)} + \tilde{\beta} b_{i} {\left( \nu^{(i)} \right)}^{2} \right] C_{1} + \left[ \beta_{A}^{(i, i-1)} + \tilde{\alpha} b_{i}^{2} \nu^{(i)} - \tilde{\beta} b_{i} \nu^{(i)} \sigma^{(i-1)} \right] C_{2}.\label{eq:order2_recur}
\end{align}

\noindent
Comparing the commutators, one can read off the recursive formula for both the $\alpha_{B}^{(i, i)}$ and $\beta_{A}^{(i, i)}$ coefficients.
Similarly applying the BCH formula for stage $A$, one obtains a symmetric expression (in parameters $a_{i}$, $b_{i}$ and coefficients $\nu^{(i)}$, $\sigma^{(i)}$ or $\alpha$, $\beta$) to Eq.~\eqref{eq:order2_recur},

\vspace{0.1cm}

\begin{equation}
  \begin{aligned}
	  \alpha_{A}^{(i, i-1)} C_{1} + \beta_{A}^{(i, i-1)} C_{2} &= \left[ \alpha_{B}^{(i-1, i-1)} + \tilde{\alpha} a_{i}^{2} \sigma^{(i-1)} - \tilde{\beta} a_{i} \nu^{(i-1)} \sigma^{(i-1)} \right] C_{1} + \\
	  &\hspace{0.5cm} + \left[ \beta_{B}^{(i-1, i-1)} + \tilde{\beta} a_{i} \left( \sigma^{(i-1)} \right)^{2} \right] C_{2},\label{eq:alpha_beta_rec}
  \end{aligned}
\end{equation}

\vspace{0.1cm}

\noindent
which shows the power of this symmetry.
With this symmetry the derivation becomes half as complicated, and higher order recursive formulae can be derived with less effort.
These recursive formulae also show a polynomial structure in the scheme parameters $a_{i}$ and $b_{i}$.
As with higher order bases, the recursive formulae become unreadable, so we do not write them here explicitly, but they are also collected in our notebook on~\cite{markomalezic_2026_18347430}.

With the coefficients within reach, we now have everything to construct a scheme at a desired order $n$.
Our goal, however is to find the maximally efficient Trotterizations, which means we need to reduce the contributions coming from coefficients in the leading-order error $\mathcal{O}(h^{n+1})$.
For example, at order $n = 2$ the contributions to the leading-order error come from $O_{3}$, which is composed of the commutators $C_{1}$, $C_{2}$ and their respective coefficients $\alpha$, $\beta$ (see Eq.~\eqref{eq:O3}).
While the commutators $C_{k}$ remain dependent on the operators $A$ and $B$ (or the model Hamiltonian $H$), it is possible to minimize the coefficients $\alpha$, $\beta$, since we know their structure according to the formulae~\eqref{eq:order2_recur}.
A similar argument works for order $n = 4$ with the $\gamma_{k}$ coefficients and so on at higher orders.
With this in mind, we define the error function per order $\textrm{Err}_{n}$ as in Ref.~\cite{OMELYAN2003272}, which is simply the Euclidean norm,

\begin{eqnarray}
	\textrm{Err}_{2} (a_{i}, b_{i}) = \sqrt{|\alpha|^{2} + |\beta|^{2}}, \quad \textrm{Err}_{4} (a_{i}, b_{i}) = \sqrt{\sum\nolimits_{k}^{6} |\gamma_{k}|^{2}},\label{eq:order_error}
\end{eqnarray}

\noindent
and similar for higher orders.
We emphasize that the error functions are with respect to the scheme parameters $a_{i}$ and $b_{i}$.
This definition reproduces the exact error if all the operators $C_i$, $D_i$, and so on are orthonormal.
In general, the operators are neither orthogonal, not do they contribute equally to the error function.
Thus, these theoretical error functions are only an approximation of the true error.
To the best of our knowledge, no better generally applicable approximation is available which is why we will stick to the errors $\textrm{Err}_{n}$.

We summarise the number of operators (i.e.\ the number of constraints) up to order $n\le 10$ in \Cref{tab:orders}.
Therein we also list the respective ranges of cycles $q$ and parameters $q+1$ for every order.
Together with the numbers of constraints and free parameters, we can evaluate which schemes with $q$ cycles have free parameters to optimize and how many of them.
It turns out that there exist regions with valid order $n \geq 6$ schemes, even though there are no free parameters.
To our knowledge the minimal no.\ cycles $q_{\textrm{min}}$ needed for an order $n = 12$ scheme has not been derived, however a simple observation of the scaling yields the pattern $q_{\textrm{min}} = 2^{\nicefrac{n}{2}} - 1$.
If this conjecture holds, then all order $n \geq 10$ schemes come without free parameters for optimization.

\begin{table}[h!]
	\centering
	\caption{The table collects how the number of constraints changes with the order $n$ and how this affects the number of free parameters a scheme has at a given number of cycles $q$ (the number of all scheme parameters goes as $q+1$).
		Order $n = 2$ and $n = 4$ both have fewer constraints, which allows us to optimize the free parameters.
		Orders $n \geq 6$ have a region of cycles $q$, where the no.\ of constraints is larger than the no.\ of scheme parameters, but valid solutions of desired order can still be found.
    Based on the scaling observation of the minimal cycles $q_{\textrm{min}} = 2^{\nicefrac{n}{2}} - 1$, we conjecture that no free parameter region exists for order $n \geq 10$.
    Information at which cycles $q$ a higher order is possible can be found inside Table 1 of~\cite{BLANES2002313}, which is obtained by following Yoshida's method~\cite{YOSHIDA1990262}.
    Parentheses denote ranges of integer values and the `*' values are only conjectured to be true.}
	\begin{tabular}{lccccccc}
		\toprule
		\multicolumn{1}{l}{Order $n$}        & \multicolumn{1}{c|}{2}      & \multicolumn{1}{c|}{4}      &                                  \multicolumn{2}{c|}{6}                                     & \multicolumn{2}{c|}{8}        &     10    \\
		\multicolumn{1}{l}{No.\ constraints} & \multicolumn{1}{c|}{2}      & \multicolumn{1}{c|}{4}      &                                  \multicolumn{2}{c|}{10}                                    & \multicolumn{2}{c|}{28}       &     84    \\
		\midrule
		\multicolumn{1}{l}{Cycles $q$}       & \multicolumn{1}{c|}{[1, 2]} & \multicolumn{1}{c|}{[3, 6]} & \multicolumn{1}{c|}{\enspace[7, 8]\enspace} & \multicolumn{1}{c|}{[9, 14]}  & \multicolumn{1}{c|}{[15, 26]} & \multicolumn{1}{c|}{[27, 30]} & [31, 62*] \\
		\multicolumn{1}{l}{No.\ parameters}  & \multicolumn{1}{c|}{[2, 3]} & \multicolumn{1}{c|}{[4, 7]} & \multicolumn{1}{c|}{\enspace[8, 9]\enspace} & \multicolumn{1}{c|}{[10, 15]} & \multicolumn{1}{c|}{[16, 27]} & \multicolumn{1}{c|}{[28, 31]} & [32, 63*] \\
		\midrule
    \multicolumn{1}{l}{Free parameters}  & \multicolumn{1}{c|}{[0, 1]} & \multicolumn{1}{c|}{[0, 3]} & \multicolumn{1}{c|}{0}      & \multicolumn{1}{c|}{[0, 5]}   & \multicolumn{1}{c|}{0}        & \multicolumn{1}{c|}{[0, 3]}   &     0     \\
		\bottomrule
	\end{tabular}\label{tab:orders}
\end{table}

\subsection{Numerical optimization}\label{sec:optim}

\begin{figure}[h]
	\begin{center}
    \resizebox{0.405\textwidth}{!}{%
		\input{Figures/Manifold_q2.pgf}
	  }
    \resizebox{0.585\textwidth}{!}{%
		\input{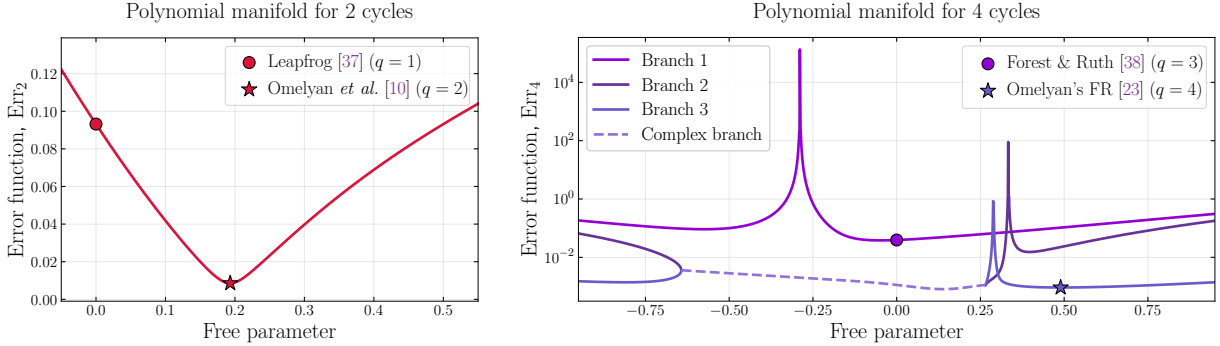}
	  }
	\end{center}
	\caption{Error manifolds of $2^{\textrm{nd}}$ order schemes at $q = 2$ cycles (left) and $4^{\textrm{th}}$ order schemes at $q = 4$ cycles (right).
		In both cases there is one real free parameter.
		The error function for 2 cycles is a simple one, with a single minimum, which is not hard to minimize.
    We plot it as a star, as well as the Leapfrog scheme, which can be found at null free parameter.
		The picture is more complicated at $q = 4$ cycles, where one finds 3 branches, two of them merging into a complex-valued parameter region.
		This manifold has 6 minima in total, which is already harder to optimize than at order $n = 2$.
    We plot the global minimum again, and present the scheme by Forest \& Ruth on the real branch, where the value of the free parameter reaches zero.
    More manifold visualizations at $q = 5$ and $q = 6$ cycles are available in our repository~\cite{markomalezic_2026_18347430}.
		We obtained the plots by solving the constraints and computing the error functions with respect to the free parameter.}\label{fig:manifolds}
\end{figure}

\noindent
From \Cref{tab:orders} we can deduce that there are cycles $q$, which allow free parameters in a Trotter-Suzuki decomposition.
Due to the polynomial nature of the coefficients, we can think of the error functions $\textrm{Err}_{n} (a_{i}, b_{i})$ as polynomial manifolds, which can be minimized in terms of the free parameters in order to find their minima.
The global minimum would then in principle provide the most efficient scheme at a given number of cycles $q$.
As the parameter number grows, the error manifolds become multidimensional and quite intricate due to the constraints one has to satisfy at increased orders.
We show examples of two such manifolds at orders $n = 2, 4$ in \Cref{fig:manifolds}, where we have a single free parameter.
The $2^{\textrm{nd}}$ order manifold at $q = 2$ cycles (left) is a simple function with a single minimum, which is also the one that Omelyan \textit{et al.}~\cite{OMELYAN2003272} give as their result.
Only two cycles above at $q = 4$ and order $n = 4$ (right) we find a much more complicated error manifold, which has three branch solutions, each with 2 minima.
The upper branch in violet is completely real, while there is a region of complex-valued solutions where the other two branches merge.
We collect manifolds at $q = 5$ and $q = 6$ in our repository~\cite{markomalezic_2026_18347430}, which have more free parameters and are thus harder to visualize.
However, they provide a good idea of how the manifold complexity scales with higher cycles, because there exist many more branch solutions.
It is not hard to imagine that a manifold at higher order will have even more branches and minima, therefore we need a good way to minimize it.

Having derived the scheme coefficients $\alpha, \beta, \gamma_{k} \ldots$ and defined the error functions $\textrm{Err}_{n}$ in terms of scheme parameters, we have effectively reduced the scheme construction to a minimization problem.
The general idea of our minimization procedure is as follows (see \Cref{sec:framework_details} for the technical details).
Since we do not solve the order constraints exactly, we must build a global $\chi^{2}$ function from theoretical errors $\textrm{Err}_{k}$ defined in equation~\eqref{eq:order_error} up to a desired order $n$,

\begin{equation}
  \chi^{2} = \sum_{k = 2, 4, \ldots}^{n} w_{k} \textrm{Err}_{k}^{2},\label{eq:chi2_minim}
\end{equation}

\noindent
where we scale the specific errors $\textrm{Err}_{k}$ by their respective weights $w_{k}$, which need to be tuned empirically (details on this can be found in \Cref{sec:framework_details}).

In regions with cycles, which permit free parameter optimization (see \Cref{tab:orders}), we first minimize the full $\chi^{2}$~\eqref{eq:chi2_minim}.
In the second step, we then impose the constraints by minimizing the $\chi^{2}$ without the leading-order error $\textrm{Err}_{n}$, or effectively by using a weight of $w_{n} = 0$.
We should note that the computation of the Hessian during the minimization plays some role in imposing the constraints.
After the first step the Hessian is kept `frozen', meaning that it does not update in the second step, which is believed to help with the effectiveness of the solutions.
Without the freezing, steps in the now unconstrained directions might be overly large.
Thus, the freezing guarantees to stay close to the minimal leading-order error while converging to an exact solution of order $n$.
For regions without free parameters, we only need to impose the constraints.
We can impose the constraints of $\textrm{Err}_{k} = 0$ only to some numerical precision, which is 18 significant digits in our simulations.
If desired, this can be improved within the framework, without much effort.

For the minimizer we decided to use the Levenberg-Marquardt algorithm~\cite{Levenberg:1944,Marquardt:1963}, chosen for its fast convergence, which is crucial in a high-dimensional parameter space.
The algorithm, however, has a tendency to get stuck in local minima, of which there are many.
To overcome this, we employed an approach, where we randomly sample initial parameters from a normal distribution and hope to cover the parameter space well enough to find the global minimum.
For the mean of the distribution we chose to use equal ramp parameters, i.e.\ $c_{i} = d_{i} = 1 / q$, which can be transformed to the stage parameters in the actual implementation.
This choice turns out to be vital, in the sense that most practically efficient schemes are found close to this origin point, which we discuss further in \Cref{sec:uniformity}.
The standard deviation is a hyperparameter in the range $\sigma \in [0.5, 2.0]$, which we tune to achieve good coverage of the parameter space without straying too far from the origin.
The routines for minimization and scheme evaluation as we have described them are also available in our repository~\cite{markomalezic_2026_18347430}.

We remark that, since we are only dealing with polynomials, methods like homotopy continuation would allow us to find all the (complex-valued) solutions.
Unfortunately, such an exact approach with guarantees becomes quickly infeasible.
At order $n$ Trotter schemes the corresponding $\chi^2$ function is of order $2n$ and has $2n$ complex roots in each variable.
Since we have the number of cycles $q$ free variables, the total number of roots scales as $(2n)^q$.
While there might still be some hope for lower orders $n\le4$, the resources needed from order $n\ge 6$ on are beyond hope.

It is interesting to observe that, starting from order $n\ge6$, solutions are possible with fewer parameters than constraints.
These ``accidental'' solutions of underdetermined systems of equations are not elevated to 1D manifolds even when the number of cycles is increased $q\mapsto q+1$.
This makes the search for solutions at high $q$ even more difficult.
We expect that for all large orders $n\ge 10$ only accidental solutions exist and no free-parameter optimization is possible.

%% file: Figures/Manifold_q2.pgf
%% Creator: Matplotlib, PGF backend
%%
%% To include the figure in your LaTeX document, write
%%   \input{<filename>.pgf}
%%
%% Make sure the required packages are loaded in your preamble
%%   \usepackage{pgf}
%%
%% Also ensure that all the required font packages are loaded; for instance,
%% the lmodern package is sometimes necessary when using math font.
%%   \usepackage{lmodern}
%%
%% Figures using additional raster images can only be included by \input if
%% they are in the same directory as the main LaTeX file. For loading figures
%% from other directories you can use the `import` package
%%   \usepackage{import}
%%
%% and then include the figures with
%%   \import{<path to file>}{<filename>.pgf}
%%
%% Matplotlib used the following preamble
%%
\begingroup%
\makeatletter%
\begin{pgfpicture}%
\pgfpathrectangle{\pgfpointorigin}{\pgfqpoint{6.500000in}{4.800000in}}%
\pgfusepath{use as bounding box, clip}%
\begin{pgfscope}%
\pgfsetbuttcap%
\pgfsetmiterjoin%
\definecolor{currentfill}{rgb}{1.000000,1.000000,1.000000}%
\pgfsetfillcolor{currentfill}%
\pgfsetlinewidth{0.000000pt}%
\definecolor{currentstroke}{rgb}{1.000000,1.000000,1.000000}%
\pgfsetstrokecolor{currentstroke}%
\pgfsetdash{}{0pt}%
\pgfpathmoveto{\pgfqpoint{0.000000in}{0.000000in}}%
\pgfpathlineto{\pgfqpoint{6.500000in}{0.000000in}}%
\pgfpathlineto{\pgfqpoint{6.500000in}{4.800000in}}%
\pgfpathlineto{\pgfqpoint{0.000000in}{4.800000in}}%
\pgfpathlineto{\pgfqpoint{0.000000in}{0.000000in}}%
\pgfpathclose%
\pgfusepath{fill}%
\end{pgfscope}%
\begin{pgfscope}%
\pgfsetbuttcap%
\pgfsetmiterjoin%
\definecolor{currentfill}{rgb}{1.000000,1.000000,1.000000}%
\pgfsetfillcolor{currentfill}%
\pgfsetlinewidth{0.000000pt}%
\definecolor{currentstroke}{rgb}{0.000000,0.000000,0.000000}%
\pgfsetstrokecolor{currentstroke}%
\pgfsetstrokeopacity{0.000000}%
\pgfsetdash{}{0pt}%
\pgfpathmoveto{\pgfqpoint{0.888378in}{0.718012in}}%
\pgfpathlineto{\pgfqpoint{6.320000in}{0.718012in}}%
\pgfpathlineto{\pgfqpoint{6.320000in}{4.151828in}}%
\pgfpathlineto{\pgfqpoint{0.888378in}{4.151828in}}%
\pgfpathlineto{\pgfqpoint{0.888378in}{0.718012in}}%
\pgfpathclose%
\pgfusepath{fill}%
\end{pgfscope}%
\begin{pgfscope}%
\pgfpathrectangle{\pgfqpoint{0.888378in}{0.718012in}}{\pgfqpoint{5.431622in}{3.433817in}}%
\pgfusepath{clip}%
\pgfsetrectcap%
\pgfsetroundjoin%
\pgfsetlinewidth{0.501875pt}%
\definecolor{currentstroke}{rgb}{0.690196,0.690196,0.690196}%
\pgfsetstrokecolor{currentstroke}%
\pgfsetstrokeopacity{0.300000}%
\pgfsetdash{}{0pt}%
\pgfpathmoveto{\pgfqpoint{1.341013in}{0.718012in}}%
\pgfpathlineto{\pgfqpoint{1.341013in}{4.151828in}}%
\pgfusepath{stroke}%
\end{pgfscope}%
\begin{pgfscope}%
\pgfsetbuttcap%
\pgfsetroundjoin%
\definecolor{currentfill}{rgb}{0.000000,0.000000,0.000000}%
\pgfsetfillcolor{currentfill}%
\pgfsetlinewidth{1.003750pt}%
\definecolor{currentstroke}{rgb}{0.000000,0.000000,0.000000}%
\pgfsetstrokecolor{currentstroke}%
\pgfsetdash{}{0pt}%
\pgfsys@defobject{currentmarker}{\pgfqpoint{0.000000in}{0.000000in}}{\pgfqpoint{0.000000in}{0.055556in}}{%
\pgfpathmoveto{\pgfqpoint{0.000000in}{0.000000in}}%
\pgfpathlineto{\pgfqpoint{0.000000in}{0.055556in}}%
\pgfusepath{stroke,fill}%
}%
\begin{pgfscope}%
\pgfsys@transformshift{1.341013in}{0.718012in}%
\pgfsys@useobject{currentmarker}{}%
\end{pgfscope}%
\end{pgfscope}%
\begin{pgfscope}%
\pgfsetbuttcap%
\pgfsetroundjoin%
\definecolor{currentfill}{rgb}{0.000000,0.000000,0.000000}%
\pgfsetfillcolor{currentfill}%
\pgfsetlinewidth{1.003750pt}%
\definecolor{currentstroke}{rgb}{0.000000,0.000000,0.000000}%
\pgfsetstrokecolor{currentstroke}%
\pgfsetdash{}{0pt}%
\pgfsys@defobject{currentmarker}{\pgfqpoint{0.000000in}{-0.055556in}}{\pgfqpoint{0.000000in}{0.000000in}}{%
\pgfpathmoveto{\pgfqpoint{0.000000in}{0.000000in}}%
\pgfpathlineto{\pgfqpoint{0.000000in}{-0.055556in}}%
\pgfusepath{stroke,fill}%
}%
\begin{pgfscope}%
\pgfsys@transformshift{1.341013in}{4.151828in}%
\pgfsys@useobject{currentmarker}{}%
\end{pgfscope}%
\end{pgfscope}%
\begin{pgfscope}%
\definecolor{textcolor}{rgb}{0.000000,0.000000,0.000000}%
\pgfsetstrokecolor{textcolor}%
\pgfsetfillcolor{textcolor}%
\pgftext[x=1.341013in,y=0.669401in,,top]{\color{textcolor}\sffamily\fontsize{15.000000}{18.000000}\selectfont \(\displaystyle {0.0}\)}%
\end{pgfscope}%
\begin{pgfscope}%
\pgfpathrectangle{\pgfqpoint{0.888378in}{0.718012in}}{\pgfqpoint{5.431622in}{3.433817in}}%
\pgfusepath{clip}%
\pgfsetrectcap%
\pgfsetroundjoin%
\pgfsetlinewidth{0.501875pt}%
\definecolor{currentstroke}{rgb}{0.690196,0.690196,0.690196}%
\pgfsetstrokecolor{currentstroke}%
\pgfsetstrokeopacity{0.300000}%
\pgfsetdash{}{0pt}%
\pgfpathmoveto{\pgfqpoint{2.246284in}{0.718012in}}%
\pgfpathlineto{\pgfqpoint{2.246284in}{4.151828in}}%
\pgfusepath{stroke}%
\end{pgfscope}%
\begin{pgfscope}%
\pgfsetbuttcap%
\pgfsetroundjoin%
\definecolor{currentfill}{rgb}{0.000000,0.000000,0.000000}%
\pgfsetfillcolor{currentfill}%
\pgfsetlinewidth{1.003750pt}%
\definecolor{currentstroke}{rgb}{0.000000,0.000000,0.000000}%
\pgfsetstrokecolor{currentstroke}%
\pgfsetdash{}{0pt}%
\pgfsys@defobject{currentmarker}{\pgfqpoint{0.000000in}{0.000000in}}{\pgfqpoint{0.000000in}{0.055556in}}{%
\pgfpathmoveto{\pgfqpoint{0.000000in}{0.000000in}}%
\pgfpathlineto{\pgfqpoint{0.000000in}{0.055556in}}%
\pgfusepath{stroke,fill}%
}%
\begin{pgfscope}%
\pgfsys@transformshift{2.246284in}{0.718012in}%
\pgfsys@useobject{currentmarker}{}%
\end{pgfscope}%
\end{pgfscope}%
\begin{pgfscope}%
\pgfsetbuttcap%
\pgfsetroundjoin%
\definecolor{currentfill}{rgb}{0.000000,0.000000,0.000000}%
\pgfsetfillcolor{currentfill}%
\pgfsetlinewidth{1.003750pt}%
\definecolor{currentstroke}{rgb}{0.000000,0.000000,0.000000}%
\pgfsetstrokecolor{currentstroke}%
\pgfsetdash{}{0pt}%
\pgfsys@defobject{currentmarker}{\pgfqpoint{0.000000in}{-0.055556in}}{\pgfqpoint{0.000000in}{0.000000in}}{%
\pgfpathmoveto{\pgfqpoint{0.000000in}{0.000000in}}%
\pgfpathlineto{\pgfqpoint{0.000000in}{-0.055556in}}%
\pgfusepath{stroke,fill}%
}%
\begin{pgfscope}%
\pgfsys@transformshift{2.246284in}{4.151828in}%
\pgfsys@useobject{currentmarker}{}%
\end{pgfscope}%
\end{pgfscope}%
\begin{pgfscope}%
\definecolor{textcolor}{rgb}{0.000000,0.000000,0.000000}%
\pgfsetstrokecolor{textcolor}%
\pgfsetfillcolor{textcolor}%
\pgftext[x=2.246284in,y=0.669401in,,top]{\color{textcolor}\sffamily\fontsize{15.000000}{18.000000}\selectfont \(\displaystyle {0.1}\)}%
\end{pgfscope}%
\begin{pgfscope}%
\pgfpathrectangle{\pgfqpoint{0.888378in}{0.718012in}}{\pgfqpoint{5.431622in}{3.433817in}}%
\pgfusepath{clip}%
\pgfsetrectcap%
\pgfsetroundjoin%
\pgfsetlinewidth{0.501875pt}%
\definecolor{currentstroke}{rgb}{0.690196,0.690196,0.690196}%
\pgfsetstrokecolor{currentstroke}%
\pgfsetstrokeopacity{0.300000}%
\pgfsetdash{}{0pt}%
\pgfpathmoveto{\pgfqpoint{3.151554in}{0.718012in}}%
\pgfpathlineto{\pgfqpoint{3.151554in}{4.151828in}}%
\pgfusepath{stroke}%
\end{pgfscope}%
\begin{pgfscope}%
\pgfsetbuttcap%
\pgfsetroundjoin%
\definecolor{currentfill}{rgb}{0.000000,0.000000,0.000000}%
\pgfsetfillcolor{currentfill}%
\pgfsetlinewidth{1.003750pt}%
\definecolor{currentstroke}{rgb}{0.000000,0.000000,0.000000}%
\pgfsetstrokecolor{currentstroke}%
\pgfsetdash{}{0pt}%
\pgfsys@defobject{currentmarker}{\pgfqpoint{0.000000in}{0.000000in}}{\pgfqpoint{0.000000in}{0.055556in}}{%
\pgfpathmoveto{\pgfqpoint{0.000000in}{0.000000in}}%
\pgfpathlineto{\pgfqpoint{0.000000in}{0.055556in}}%
\pgfusepath{stroke,fill}%
}%
\begin{pgfscope}%
\pgfsys@transformshift{3.151554in}{0.718012in}%
\pgfsys@useobject{currentmarker}{}%
\end{pgfscope}%
\end{pgfscope}%
\begin{pgfscope}%
\pgfsetbuttcap%
\pgfsetroundjoin%
\definecolor{currentfill}{rgb}{0.000000,0.000000,0.000000}%
\pgfsetfillcolor{currentfill}%
\pgfsetlinewidth{1.003750pt}%
\definecolor{currentstroke}{rgb}{0.000000,0.000000,0.000000}%
\pgfsetstrokecolor{currentstroke}%
\pgfsetdash{}{0pt}%
\pgfsys@defobject{currentmarker}{\pgfqpoint{0.000000in}{-0.055556in}}{\pgfqpoint{0.000000in}{0.000000in}}{%
\pgfpathmoveto{\pgfqpoint{0.000000in}{0.000000in}}%
\pgfpathlineto{\pgfqpoint{0.000000in}{-0.055556in}}%
\pgfusepath{stroke,fill}%
}%
\begin{pgfscope}%
\pgfsys@transformshift{3.151554in}{4.151828in}%
\pgfsys@useobject{currentmarker}{}%
\end{pgfscope}%
\end{pgfscope}%
\begin{pgfscope}%
\definecolor{textcolor}{rgb}{0.000000,0.000000,0.000000}%
\pgfsetstrokecolor{textcolor}%
\pgfsetfillcolor{textcolor}%
\pgftext[x=3.151554in,y=0.669401in,,top]{\color{textcolor}\sffamily\fontsize{15.000000}{18.000000}\selectfont \(\displaystyle {0.2}\)}%
\end{pgfscope}%
\begin{pgfscope}%
\pgfpathrectangle{\pgfqpoint{0.888378in}{0.718012in}}{\pgfqpoint{5.431622in}{3.433817in}}%
\pgfusepath{clip}%
\pgfsetrectcap%
\pgfsetroundjoin%
\pgfsetlinewidth{0.501875pt}%
\definecolor{currentstroke}{rgb}{0.690196,0.690196,0.690196}%
\pgfsetstrokecolor{currentstroke}%
\pgfsetstrokeopacity{0.300000}%
\pgfsetdash{}{0pt}%
\pgfpathmoveto{\pgfqpoint{4.056824in}{0.718012in}}%
\pgfpathlineto{\pgfqpoint{4.056824in}{4.151828in}}%
\pgfusepath{stroke}%
\end{pgfscope}%
\begin{pgfscope}%
\pgfsetbuttcap%
\pgfsetroundjoin%
\definecolor{currentfill}{rgb}{0.000000,0.000000,0.000000}%
\pgfsetfillcolor{currentfill}%
\pgfsetlinewidth{1.003750pt}%
\definecolor{currentstroke}{rgb}{0.000000,0.000000,0.000000}%
\pgfsetstrokecolor{currentstroke}%
\pgfsetdash{}{0pt}%
\pgfsys@defobject{currentmarker}{\pgfqpoint{0.000000in}{0.000000in}}{\pgfqpoint{0.000000in}{0.055556in}}{%
\pgfpathmoveto{\pgfqpoint{0.000000in}{0.000000in}}%
\pgfpathlineto{\pgfqpoint{0.000000in}{0.055556in}}%
\pgfusepath{stroke,fill}%
}%
\begin{pgfscope}%
\pgfsys@transformshift{4.056824in}{0.718012in}%
\pgfsys@useobject{currentmarker}{}%
\end{pgfscope}%
\end{pgfscope}%
\begin{pgfscope}%
\pgfsetbuttcap%
\pgfsetroundjoin%
\definecolor{currentfill}{rgb}{0.000000,0.000000,0.000000}%
\pgfsetfillcolor{currentfill}%
\pgfsetlinewidth{1.003750pt}%
\definecolor{currentstroke}{rgb}{0.000000,0.000000,0.000000}%
\pgfsetstrokecolor{currentstroke}%
\pgfsetdash{}{0pt}%
\pgfsys@defobject{currentmarker}{\pgfqpoint{0.000000in}{-0.055556in}}{\pgfqpoint{0.000000in}{0.000000in}}{%
\pgfpathmoveto{\pgfqpoint{0.000000in}{0.000000in}}%
\pgfpathlineto{\pgfqpoint{0.000000in}{-0.055556in}}%
\pgfusepath{stroke,fill}%
}%
\begin{pgfscope}%
\pgfsys@transformshift{4.056824in}{4.151828in}%
\pgfsys@useobject{currentmarker}{}%
\end{pgfscope}%
\end{pgfscope}%
\begin{pgfscope}%
\definecolor{textcolor}{rgb}{0.000000,0.000000,0.000000}%
\pgfsetstrokecolor{textcolor}%
\pgfsetfillcolor{textcolor}%
\pgftext[x=4.056824in,y=0.669401in,,top]{\color{textcolor}\sffamily\fontsize{15.000000}{18.000000}\selectfont \(\displaystyle {0.3}\)}%
\end{pgfscope}%
\begin{pgfscope}%
\pgfpathrectangle{\pgfqpoint{0.888378in}{0.718012in}}{\pgfqpoint{5.431622in}{3.433817in}}%
\pgfusepath{clip}%
\pgfsetrectcap%
\pgfsetroundjoin%
\pgfsetlinewidth{0.501875pt}%
\definecolor{currentstroke}{rgb}{0.690196,0.690196,0.690196}%
\pgfsetstrokecolor{currentstroke}%
\pgfsetstrokeopacity{0.300000}%
\pgfsetdash{}{0pt}%
\pgfpathmoveto{\pgfqpoint{4.962095in}{0.718012in}}%
\pgfpathlineto{\pgfqpoint{4.962095in}{4.151828in}}%
\pgfusepath{stroke}%
\end{pgfscope}%
\begin{pgfscope}%
\pgfsetbuttcap%
\pgfsetroundjoin%
\definecolor{currentfill}{rgb}{0.000000,0.000000,0.000000}%
\pgfsetfillcolor{currentfill}%
\pgfsetlinewidth{1.003750pt}%
\definecolor{currentstroke}{rgb}{0.000000,0.000000,0.000000}%
\pgfsetstrokecolor{currentstroke}%
\pgfsetdash{}{0pt}%
\pgfsys@defobject{currentmarker}{\pgfqpoint{0.000000in}{0.000000in}}{\pgfqpoint{0.000000in}{0.055556in}}{%
\pgfpathmoveto{\pgfqpoint{0.000000in}{0.000000in}}%
\pgfpathlineto{\pgfqpoint{0.000000in}{0.055556in}}%
\pgfusepath{stroke,fill}%
}%
\begin{pgfscope}%
\pgfsys@transformshift{4.962095in}{0.718012in}%
\pgfsys@useobject{currentmarker}{}%
\end{pgfscope}%
\end{pgfscope}%
\begin{pgfscope}%
\pgfsetbuttcap%
\pgfsetroundjoin%
\definecolor{currentfill}{rgb}{0.000000,0.000000,0.000000}%
\pgfsetfillcolor{currentfill}%
\pgfsetlinewidth{1.003750pt}%
\definecolor{currentstroke}{rgb}{0.000000,0.000000,0.000000}%
\pgfsetstrokecolor{currentstroke}%
\pgfsetdash{}{0pt}%
\pgfsys@defobject{currentmarker}{\pgfqpoint{0.000000in}{-0.055556in}}{\pgfqpoint{0.000000in}{0.000000in}}{%
\pgfpathmoveto{\pgfqpoint{0.000000in}{0.000000in}}%
\pgfpathlineto{\pgfqpoint{0.000000in}{-0.055556in}}%
\pgfusepath{stroke,fill}%
}%
\begin{pgfscope}%
\pgfsys@transformshift{4.962095in}{4.151828in}%
\pgfsys@useobject{currentmarker}{}%
\end{pgfscope}%
\end{pgfscope}%
\begin{pgfscope}%
\definecolor{textcolor}{rgb}{0.000000,0.000000,0.000000}%
\pgfsetstrokecolor{textcolor}%
\pgfsetfillcolor{textcolor}%
\pgftext[x=4.962095in,y=0.669401in,,top]{\color{textcolor}\sffamily\fontsize{15.000000}{18.000000}\selectfont \(\displaystyle {0.4}\)}%
\end{pgfscope}%
\begin{pgfscope}%
\pgfpathrectangle{\pgfqpoint{0.888378in}{0.718012in}}{\pgfqpoint{5.431622in}{3.433817in}}%
\pgfusepath{clip}%
\pgfsetrectcap%
\pgfsetroundjoin%
\pgfsetlinewidth{0.501875pt}%
\definecolor{currentstroke}{rgb}{0.690196,0.690196,0.690196}%
\pgfsetstrokecolor{currentstroke}%
\pgfsetstrokeopacity{0.300000}%
\pgfsetdash{}{0pt}%
\pgfpathmoveto{\pgfqpoint{5.867365in}{0.718012in}}%
\pgfpathlineto{\pgfqpoint{5.867365in}{4.151828in}}%
\pgfusepath{stroke}%
\end{pgfscope}%
\begin{pgfscope}%
\pgfsetbuttcap%
\pgfsetroundjoin%
\definecolor{currentfill}{rgb}{0.000000,0.000000,0.000000}%
\pgfsetfillcolor{currentfill}%
\pgfsetlinewidth{1.003750pt}%
\definecolor{currentstroke}{rgb}{0.000000,0.000000,0.000000}%
\pgfsetstrokecolor{currentstroke}%
\pgfsetdash{}{0pt}%
\pgfsys@defobject{currentmarker}{\pgfqpoint{0.000000in}{0.000000in}}{\pgfqpoint{0.000000in}{0.055556in}}{%
\pgfpathmoveto{\pgfqpoint{0.000000in}{0.000000in}}%
\pgfpathlineto{\pgfqpoint{0.000000in}{0.055556in}}%
\pgfusepath{stroke,fill}%
}%
\begin{pgfscope}%
\pgfsys@transformshift{5.867365in}{0.718012in}%
\pgfsys@useobject{currentmarker}{}%
\end{pgfscope}%
\end{pgfscope}%
\begin{pgfscope}%
\pgfsetbuttcap%
\pgfsetroundjoin%
\definecolor{currentfill}{rgb}{0.000000,0.000000,0.000000}%
\pgfsetfillcolor{currentfill}%
\pgfsetlinewidth{1.003750pt}%
\definecolor{currentstroke}{rgb}{0.000000,0.000000,0.000000}%
\pgfsetstrokecolor{currentstroke}%
\pgfsetdash{}{0pt}%
\pgfsys@defobject{currentmarker}{\pgfqpoint{0.000000in}{-0.055556in}}{\pgfqpoint{0.000000in}{0.000000in}}{%
\pgfpathmoveto{\pgfqpoint{0.000000in}{0.000000in}}%
\pgfpathlineto{\pgfqpoint{0.000000in}{-0.055556in}}%
\pgfusepath{stroke,fill}%
}%
\begin{pgfscope}%
\pgfsys@transformshift{5.867365in}{4.151828in}%
\pgfsys@useobject{currentmarker}{}%
\end{pgfscope}%
\end{pgfscope}%
\begin{pgfscope}%
\definecolor{textcolor}{rgb}{0.000000,0.000000,0.000000}%
\pgfsetstrokecolor{textcolor}%
\pgfsetfillcolor{textcolor}%
\pgftext[x=5.867365in,y=0.669401in,,top]{\color{textcolor}\sffamily\fontsize{15.000000}{18.000000}\selectfont \(\displaystyle {0.5}\)}%
\end{pgfscope}%
\begin{pgfscope}%
\definecolor{textcolor}{rgb}{0.000000,0.000000,0.000000}%
\pgfsetstrokecolor{textcolor}%
\pgfsetfillcolor{textcolor}%
\pgftext[x=3.604189in,y=0.436068in,,top]{\color{textcolor}\sffamily\fontsize{22.000000}{26.400000}\selectfont \(\displaystyle \textrm{Free parameter}\)}%
\end{pgfscope}%
\begin{pgfscope}%
\pgfpathrectangle{\pgfqpoint{0.888378in}{0.718012in}}{\pgfqpoint{5.431622in}{3.433817in}}%
\pgfusepath{clip}%
\pgfsetrectcap%
\pgfsetroundjoin%
\pgfsetlinewidth{0.501875pt}%
\definecolor{currentstroke}{rgb}{0.690196,0.690196,0.690196}%
\pgfsetstrokecolor{currentstroke}%
\pgfsetstrokeopacity{0.300000}%
\pgfsetdash{}{0pt}%
\pgfpathmoveto{\pgfqpoint{0.888378in}{0.742539in}}%
\pgfpathlineto{\pgfqpoint{6.320000in}{0.742539in}}%
\pgfusepath{stroke}%
\end{pgfscope}%
\begin{pgfscope}%
\pgfsetbuttcap%
\pgfsetroundjoin%
\definecolor{currentfill}{rgb}{0.000000,0.000000,0.000000}%
\pgfsetfillcolor{currentfill}%
\pgfsetlinewidth{1.003750pt}%
\definecolor{currentstroke}{rgb}{0.000000,0.000000,0.000000}%
\pgfsetstrokecolor{currentstroke}%
\pgfsetdash{}{0pt}%
\pgfsys@defobject{currentmarker}{\pgfqpoint{0.000000in}{0.000000in}}{\pgfqpoint{0.055556in}{0.000000in}}{%
\pgfpathmoveto{\pgfqpoint{0.000000in}{0.000000in}}%
\pgfpathlineto{\pgfqpoint{0.055556in}{0.000000in}}%
\pgfusepath{stroke,fill}%
}%
\begin{pgfscope}%
\pgfsys@transformshift{0.888378in}{0.742539in}%
\pgfsys@useobject{currentmarker}{}%
\end{pgfscope}%
\end{pgfscope}%
\begin{pgfscope}%
\pgfsetbuttcap%
\pgfsetroundjoin%
\definecolor{currentfill}{rgb}{0.000000,0.000000,0.000000}%
\pgfsetfillcolor{currentfill}%
\pgfsetlinewidth{1.003750pt}%
\definecolor{currentstroke}{rgb}{0.000000,0.000000,0.000000}%
\pgfsetstrokecolor{currentstroke}%
\pgfsetdash{}{0pt}%
\pgfsys@defobject{currentmarker}{\pgfqpoint{-0.055556in}{0.000000in}}{\pgfqpoint{-0.000000in}{0.000000in}}{%
\pgfpathmoveto{\pgfqpoint{-0.000000in}{0.000000in}}%
\pgfpathlineto{\pgfqpoint{-0.055556in}{0.000000in}}%
\pgfusepath{stroke,fill}%
}%
\begin{pgfscope}%
\pgfsys@transformshift{6.320000in}{0.742539in}%
\pgfsys@useobject{currentmarker}{}%
\end{pgfscope}%
\end{pgfscope}%
\begin{pgfscope}%
\definecolor{textcolor}{rgb}{0.000000,0.000000,0.000000}%
\pgfsetstrokecolor{textcolor}%
\pgfsetfillcolor{textcolor}%
\pgftext[x=0.491623in, y=0.673095in, left, base]{\color{textcolor}\sffamily\fontsize{15.000000}{18.000000}\selectfont \(\displaystyle {0.00}\)}%
\end{pgfscope}%
\begin{pgfscope}%
\pgfpathrectangle{\pgfqpoint{0.888378in}{0.718012in}}{\pgfqpoint{5.431622in}{3.433817in}}%
\pgfusepath{clip}%
\pgfsetrectcap%
\pgfsetroundjoin%
\pgfsetlinewidth{0.501875pt}%
\definecolor{currentstroke}{rgb}{0.690196,0.690196,0.690196}%
\pgfsetstrokecolor{currentstroke}%
\pgfsetstrokeopacity{0.300000}%
\pgfsetdash{}{0pt}%
\pgfpathmoveto{\pgfqpoint{0.888378in}{1.233084in}}%
\pgfpathlineto{\pgfqpoint{6.320000in}{1.233084in}}%
\pgfusepath{stroke}%
\end{pgfscope}%
\begin{pgfscope}%
\pgfsetbuttcap%
\pgfsetroundjoin%
\definecolor{currentfill}{rgb}{0.000000,0.000000,0.000000}%
\pgfsetfillcolor{currentfill}%
\pgfsetlinewidth{1.003750pt}%
\definecolor{currentstroke}{rgb}{0.000000,0.000000,0.000000}%
\pgfsetstrokecolor{currentstroke}%
\pgfsetdash{}{0pt}%
\pgfsys@defobject{currentmarker}{\pgfqpoint{0.000000in}{0.000000in}}{\pgfqpoint{0.055556in}{0.000000in}}{%
\pgfpathmoveto{\pgfqpoint{0.000000in}{0.000000in}}%
\pgfpathlineto{\pgfqpoint{0.055556in}{0.000000in}}%
\pgfusepath{stroke,fill}%
}%
\begin{pgfscope}%
\pgfsys@transformshift{0.888378in}{1.233084in}%
\pgfsys@useobject{currentmarker}{}%
\end{pgfscope}%
\end{pgfscope}%
\begin{pgfscope}%
\pgfsetbuttcap%
\pgfsetroundjoin%
\definecolor{currentfill}{rgb}{0.000000,0.000000,0.000000}%
\pgfsetfillcolor{currentfill}%
\pgfsetlinewidth{1.003750pt}%
\definecolor{currentstroke}{rgb}{0.000000,0.000000,0.000000}%
\pgfsetstrokecolor{currentstroke}%
\pgfsetdash{}{0pt}%
\pgfsys@defobject{currentmarker}{\pgfqpoint{-0.055556in}{0.000000in}}{\pgfqpoint{-0.000000in}{0.000000in}}{%
\pgfpathmoveto{\pgfqpoint{-0.000000in}{0.000000in}}%
\pgfpathlineto{\pgfqpoint{-0.055556in}{0.000000in}}%
\pgfusepath{stroke,fill}%
}%
\begin{pgfscope}%
\pgfsys@transformshift{6.320000in}{1.233084in}%
\pgfsys@useobject{currentmarker}{}%
\end{pgfscope}%
\end{pgfscope}%
\begin{pgfscope}%
\definecolor{textcolor}{rgb}{0.000000,0.000000,0.000000}%
\pgfsetstrokecolor{textcolor}%
\pgfsetfillcolor{textcolor}%
\pgftext[x=0.491623in, y=1.163640in, left, base]{\color{textcolor}\sffamily\fontsize{15.000000}{18.000000}\selectfont \(\displaystyle {0.02}\)}%
\end{pgfscope}%
\begin{pgfscope}%
\pgfpathrectangle{\pgfqpoint{0.888378in}{0.718012in}}{\pgfqpoint{5.431622in}{3.433817in}}%
\pgfusepath{clip}%
\pgfsetrectcap%
\pgfsetroundjoin%
\pgfsetlinewidth{0.501875pt}%
\definecolor{currentstroke}{rgb}{0.690196,0.690196,0.690196}%
\pgfsetstrokecolor{currentstroke}%
\pgfsetstrokeopacity{0.300000}%
\pgfsetdash{}{0pt}%
\pgfpathmoveto{\pgfqpoint{0.888378in}{1.723629in}}%
\pgfpathlineto{\pgfqpoint{6.320000in}{1.723629in}}%
\pgfusepath{stroke}%
\end{pgfscope}%
\begin{pgfscope}%
\pgfsetbuttcap%
\pgfsetroundjoin%
\definecolor{currentfill}{rgb}{0.000000,0.000000,0.000000}%
\pgfsetfillcolor{currentfill}%
\pgfsetlinewidth{1.003750pt}%
\definecolor{currentstroke}{rgb}{0.000000,0.000000,0.000000}%
\pgfsetstrokecolor{currentstroke}%
\pgfsetdash{}{0pt}%
\pgfsys@defobject{currentmarker}{\pgfqpoint{0.000000in}{0.000000in}}{\pgfqpoint{0.055556in}{0.000000in}}{%
\pgfpathmoveto{\pgfqpoint{0.000000in}{0.000000in}}%
\pgfpathlineto{\pgfqpoint{0.055556in}{0.000000in}}%
\pgfusepath{stroke,fill}%
}%
\begin{pgfscope}%
\pgfsys@transformshift{0.888378in}{1.723629in}%
\pgfsys@useobject{currentmarker}{}%
\end{pgfscope}%
\end{pgfscope}%
\begin{pgfscope}%
\pgfsetbuttcap%
\pgfsetroundjoin%
\definecolor{currentfill}{rgb}{0.000000,0.000000,0.000000}%
\pgfsetfillcolor{currentfill}%
\pgfsetlinewidth{1.003750pt}%
\definecolor{currentstroke}{rgb}{0.000000,0.000000,0.000000}%
\pgfsetstrokecolor{currentstroke}%
\pgfsetdash{}{0pt}%
\pgfsys@defobject{currentmarker}{\pgfqpoint{-0.055556in}{0.000000in}}{\pgfqpoint{-0.000000in}{0.000000in}}{%
\pgfpathmoveto{\pgfqpoint{-0.000000in}{0.000000in}}%
\pgfpathlineto{\pgfqpoint{-0.055556in}{0.000000in}}%
\pgfusepath{stroke,fill}%
}%
\begin{pgfscope}%
\pgfsys@transformshift{6.320000in}{1.723629in}%
\pgfsys@useobject{currentmarker}{}%
\end{pgfscope}%
\end{pgfscope}%
\begin{pgfscope}%
\definecolor{textcolor}{rgb}{0.000000,0.000000,0.000000}%
\pgfsetstrokecolor{textcolor}%
\pgfsetfillcolor{textcolor}%
\pgftext[x=0.491623in, y=1.654185in, left, base]{\color{textcolor}\sffamily\fontsize{15.000000}{18.000000}\selectfont \(\displaystyle {0.04}\)}%
\end{pgfscope}%
\begin{pgfscope}%
\pgfpathrectangle{\pgfqpoint{0.888378in}{0.718012in}}{\pgfqpoint{5.431622in}{3.433817in}}%
\pgfusepath{clip}%
\pgfsetrectcap%
\pgfsetroundjoin%
\pgfsetlinewidth{0.501875pt}%
\definecolor{currentstroke}{rgb}{0.690196,0.690196,0.690196}%
\pgfsetstrokecolor{currentstroke}%
\pgfsetstrokeopacity{0.300000}%
\pgfsetdash{}{0pt}%
\pgfpathmoveto{\pgfqpoint{0.888378in}{2.214175in}}%
\pgfpathlineto{\pgfqpoint{6.320000in}{2.214175in}}%
\pgfusepath{stroke}%
\end{pgfscope}%
\begin{pgfscope}%
\pgfsetbuttcap%
\pgfsetroundjoin%
\definecolor{currentfill}{rgb}{0.000000,0.000000,0.000000}%
\pgfsetfillcolor{currentfill}%
\pgfsetlinewidth{1.003750pt}%
\definecolor{currentstroke}{rgb}{0.000000,0.000000,0.000000}%
\pgfsetstrokecolor{currentstroke}%
\pgfsetdash{}{0pt}%
\pgfsys@defobject{currentmarker}{\pgfqpoint{0.000000in}{0.000000in}}{\pgfqpoint{0.055556in}{0.000000in}}{%
\pgfpathmoveto{\pgfqpoint{0.000000in}{0.000000in}}%
\pgfpathlineto{\pgfqpoint{0.055556in}{0.000000in}}%
\pgfusepath{stroke,fill}%
}%
\begin{pgfscope}%
\pgfsys@transformshift{0.888378in}{2.214175in}%
\pgfsys@useobject{currentmarker}{}%
\end{pgfscope}%
\end{pgfscope}%
\begin{pgfscope}%
\pgfsetbuttcap%
\pgfsetroundjoin%
\definecolor{currentfill}{rgb}{0.000000,0.000000,0.000000}%
\pgfsetfillcolor{currentfill}%
\pgfsetlinewidth{1.003750pt}%
\definecolor{currentstroke}{rgb}{0.000000,0.000000,0.000000}%
\pgfsetstrokecolor{currentstroke}%
\pgfsetdash{}{0pt}%
\pgfsys@defobject{currentmarker}{\pgfqpoint{-0.055556in}{0.000000in}}{\pgfqpoint{-0.000000in}{0.000000in}}{%
\pgfpathmoveto{\pgfqpoint{-0.000000in}{0.000000in}}%
\pgfpathlineto{\pgfqpoint{-0.055556in}{0.000000in}}%
\pgfusepath{stroke,fill}%
}%
\begin{pgfscope}%
\pgfsys@transformshift{6.320000in}{2.214175in}%
\pgfsys@useobject{currentmarker}{}%
\end{pgfscope}%
\end{pgfscope}%
\begin{pgfscope}%
\definecolor{textcolor}{rgb}{0.000000,0.000000,0.000000}%
\pgfsetstrokecolor{textcolor}%
\pgfsetfillcolor{textcolor}%
\pgftext[x=0.491623in, y=2.144730in, left, base]{\color{textcolor}\sffamily\fontsize{15.000000}{18.000000}\selectfont \(\displaystyle {0.06}\)}%
\end{pgfscope}%
\begin{pgfscope}%
\pgfpathrectangle{\pgfqpoint{0.888378in}{0.718012in}}{\pgfqpoint{5.431622in}{3.433817in}}%
\pgfusepath{clip}%
\pgfsetrectcap%
\pgfsetroundjoin%
\pgfsetlinewidth{0.501875pt}%
\definecolor{currentstroke}{rgb}{0.690196,0.690196,0.690196}%
\pgfsetstrokecolor{currentstroke}%
\pgfsetstrokeopacity{0.300000}%
\pgfsetdash{}{0pt}%
\pgfpathmoveto{\pgfqpoint{0.888378in}{2.704720in}}%
\pgfpathlineto{\pgfqpoint{6.320000in}{2.704720in}}%
\pgfusepath{stroke}%
\end{pgfscope}%
\begin{pgfscope}%
\pgfsetbuttcap%
\pgfsetroundjoin%
\definecolor{currentfill}{rgb}{0.000000,0.000000,0.000000}%
\pgfsetfillcolor{currentfill}%
\pgfsetlinewidth{1.003750pt}%
\definecolor{currentstroke}{rgb}{0.000000,0.000000,0.000000}%
\pgfsetstrokecolor{currentstroke}%
\pgfsetdash{}{0pt}%
\pgfsys@defobject{currentmarker}{\pgfqpoint{0.000000in}{0.000000in}}{\pgfqpoint{0.055556in}{0.000000in}}{%
\pgfpathmoveto{\pgfqpoint{0.000000in}{0.000000in}}%
\pgfpathlineto{\pgfqpoint{0.055556in}{0.000000in}}%
\pgfusepath{stroke,fill}%
}%
\begin{pgfscope}%
\pgfsys@transformshift{0.888378in}{2.704720in}%
\pgfsys@useobject{currentmarker}{}%
\end{pgfscope}%
\end{pgfscope}%
\begin{pgfscope}%
\pgfsetbuttcap%
\pgfsetroundjoin%
\definecolor{currentfill}{rgb}{0.000000,0.000000,0.000000}%
\pgfsetfillcolor{currentfill}%
\pgfsetlinewidth{1.003750pt}%
\definecolor{currentstroke}{rgb}{0.000000,0.000000,0.000000}%
\pgfsetstrokecolor{currentstroke}%
\pgfsetdash{}{0pt}%
\pgfsys@defobject{currentmarker}{\pgfqpoint{-0.055556in}{0.000000in}}{\pgfqpoint{-0.000000in}{0.000000in}}{%
\pgfpathmoveto{\pgfqpoint{-0.000000in}{0.000000in}}%
\pgfpathlineto{\pgfqpoint{-0.055556in}{0.000000in}}%
\pgfusepath{stroke,fill}%
}%
\begin{pgfscope}%
\pgfsys@transformshift{6.320000in}{2.704720in}%
\pgfsys@useobject{currentmarker}{}%
\end{pgfscope}%
\end{pgfscope}%
\begin{pgfscope}%
\definecolor{textcolor}{rgb}{0.000000,0.000000,0.000000}%
\pgfsetstrokecolor{textcolor}%
\pgfsetfillcolor{textcolor}%
\pgftext[x=0.491623in, y=2.635275in, left, base]{\color{textcolor}\sffamily\fontsize{15.000000}{18.000000}\selectfont \(\displaystyle {0.08}\)}%
\end{pgfscope}%
\begin{pgfscope}%
\pgfpathrectangle{\pgfqpoint{0.888378in}{0.718012in}}{\pgfqpoint{5.431622in}{3.433817in}}%
\pgfusepath{clip}%
\pgfsetrectcap%
\pgfsetroundjoin%
\pgfsetlinewidth{0.501875pt}%
\definecolor{currentstroke}{rgb}{0.690196,0.690196,0.690196}%
\pgfsetstrokecolor{currentstroke}%
\pgfsetstrokeopacity{0.300000}%
\pgfsetdash{}{0pt}%
\pgfpathmoveto{\pgfqpoint{0.888378in}{3.195265in}}%
\pgfpathlineto{\pgfqpoint{6.320000in}{3.195265in}}%
\pgfusepath{stroke}%
\end{pgfscope}%
\begin{pgfscope}%
\pgfsetbuttcap%
\pgfsetroundjoin%
\definecolor{currentfill}{rgb}{0.000000,0.000000,0.000000}%
\pgfsetfillcolor{currentfill}%
\pgfsetlinewidth{1.003750pt}%
\definecolor{currentstroke}{rgb}{0.000000,0.000000,0.000000}%
\pgfsetstrokecolor{currentstroke}%
\pgfsetdash{}{0pt}%
\pgfsys@defobject{currentmarker}{\pgfqpoint{0.000000in}{0.000000in}}{\pgfqpoint{0.055556in}{0.000000in}}{%
\pgfpathmoveto{\pgfqpoint{0.000000in}{0.000000in}}%
\pgfpathlineto{\pgfqpoint{0.055556in}{0.000000in}}%
\pgfusepath{stroke,fill}%
}%
\begin{pgfscope}%
\pgfsys@transformshift{0.888378in}{3.195265in}%
\pgfsys@useobject{currentmarker}{}%
\end{pgfscope}%
\end{pgfscope}%
\begin{pgfscope}%
\pgfsetbuttcap%
\pgfsetroundjoin%
\definecolor{currentfill}{rgb}{0.000000,0.000000,0.000000}%
\pgfsetfillcolor{currentfill}%
\pgfsetlinewidth{1.003750pt}%
\definecolor{currentstroke}{rgb}{0.000000,0.000000,0.000000}%
\pgfsetstrokecolor{currentstroke}%
\pgfsetdash{}{0pt}%
\pgfsys@defobject{currentmarker}{\pgfqpoint{-0.055556in}{0.000000in}}{\pgfqpoint{-0.000000in}{0.000000in}}{%
\pgfpathmoveto{\pgfqpoint{-0.000000in}{0.000000in}}%
\pgfpathlineto{\pgfqpoint{-0.055556in}{0.000000in}}%
\pgfusepath{stroke,fill}%
}%
\begin{pgfscope}%
\pgfsys@transformshift{6.320000in}{3.195265in}%
\pgfsys@useobject{currentmarker}{}%
\end{pgfscope}%
\end{pgfscope}%
\begin{pgfscope}%
\definecolor{textcolor}{rgb}{0.000000,0.000000,0.000000}%
\pgfsetstrokecolor{textcolor}%
\pgfsetfillcolor{textcolor}%
\pgftext[x=0.491623in, y=3.125821in, left, base]{\color{textcolor}\sffamily\fontsize{15.000000}{18.000000}\selectfont \(\displaystyle {0.10}\)}%
\end{pgfscope}%
\begin{pgfscope}%
\pgfpathrectangle{\pgfqpoint{0.888378in}{0.718012in}}{\pgfqpoint{5.431622in}{3.433817in}}%
\pgfusepath{clip}%
\pgfsetrectcap%
\pgfsetroundjoin%
\pgfsetlinewidth{0.501875pt}%
\definecolor{currentstroke}{rgb}{0.690196,0.690196,0.690196}%
\pgfsetstrokecolor{currentstroke}%
\pgfsetstrokeopacity{0.300000}%
\pgfsetdash{}{0pt}%
\pgfpathmoveto{\pgfqpoint{0.888378in}{3.685810in}}%
\pgfpathlineto{\pgfqpoint{6.320000in}{3.685810in}}%
\pgfusepath{stroke}%
\end{pgfscope}%
\begin{pgfscope}%
\pgfsetbuttcap%
\pgfsetroundjoin%
\definecolor{currentfill}{rgb}{0.000000,0.000000,0.000000}%
\pgfsetfillcolor{currentfill}%
\pgfsetlinewidth{1.003750pt}%
\definecolor{currentstroke}{rgb}{0.000000,0.000000,0.000000}%
\pgfsetstrokecolor{currentstroke}%
\pgfsetdash{}{0pt}%
\pgfsys@defobject{currentmarker}{\pgfqpoint{0.000000in}{0.000000in}}{\pgfqpoint{0.055556in}{0.000000in}}{%
\pgfpathmoveto{\pgfqpoint{0.000000in}{0.000000in}}%
\pgfpathlineto{\pgfqpoint{0.055556in}{0.000000in}}%
\pgfusepath{stroke,fill}%
}%
\begin{pgfscope}%
\pgfsys@transformshift{0.888378in}{3.685810in}%
\pgfsys@useobject{currentmarker}{}%
\end{pgfscope}%
\end{pgfscope}%
\begin{pgfscope}%
\pgfsetbuttcap%
\pgfsetroundjoin%
\definecolor{currentfill}{rgb}{0.000000,0.000000,0.000000}%
\pgfsetfillcolor{currentfill}%
\pgfsetlinewidth{1.003750pt}%
\definecolor{currentstroke}{rgb}{0.000000,0.000000,0.000000}%
\pgfsetstrokecolor{currentstroke}%
\pgfsetdash{}{0pt}%
\pgfsys@defobject{currentmarker}{\pgfqpoint{-0.055556in}{0.000000in}}{\pgfqpoint{-0.000000in}{0.000000in}}{%
\pgfpathmoveto{\pgfqpoint{-0.000000in}{0.000000in}}%
\pgfpathlineto{\pgfqpoint{-0.055556in}{0.000000in}}%
\pgfusepath{stroke,fill}%
}%
\begin{pgfscope}%
\pgfsys@transformshift{6.320000in}{3.685810in}%
\pgfsys@useobject{currentmarker}{}%
\end{pgfscope}%
\end{pgfscope}%
\begin{pgfscope}%
\definecolor{textcolor}{rgb}{0.000000,0.000000,0.000000}%
\pgfsetstrokecolor{textcolor}%
\pgfsetfillcolor{textcolor}%
\pgftext[x=0.491623in, y=3.616366in, left, base]{\color{textcolor}\sffamily\fontsize{15.000000}{18.000000}\selectfont \(\displaystyle {0.12}\)}%
\end{pgfscope}%
\begin{pgfscope}%
\definecolor{textcolor}{rgb}{0.000000,0.000000,0.000000}%
\pgfsetstrokecolor{textcolor}%
\pgfsetfillcolor{textcolor}%
\pgftext[x=0.436068in,y=2.434920in,,bottom,rotate=90.000000]{\color{textcolor}\sffamily\fontsize{22.000000}{26.400000}\selectfont \(\displaystyle \textrm{Error function}\), \(\displaystyle \textrm{Err}_{2}\)}%
\end{pgfscope}%
\begin{pgfscope}%
\pgfpathrectangle{\pgfqpoint{0.888378in}{0.718012in}}{\pgfqpoint{5.431622in}{3.433817in}}%
\pgfusepath{clip}%
\pgfsetrectcap%
\pgfsetroundjoin%
\pgfsetlinewidth{2.509375pt}%
\definecolor{currentstroke}{rgb}{0.862745,0.078431,0.235294}%
\pgfsetstrokecolor{currentstroke}%
\pgfsetdash{}{0pt}%
\pgfpathmoveto{\pgfqpoint{0.888378in}{3.740743in}}%
\pgfpathlineto{\pgfqpoint{1.013430in}{3.538258in}}%
\pgfpathlineto{\pgfqpoint{1.138483in}{3.339974in}}%
\pgfpathlineto{\pgfqpoint{1.263535in}{3.145883in}}%
\pgfpathlineto{\pgfqpoint{1.388587in}{2.955979in}}%
\pgfpathlineto{\pgfqpoint{1.513640in}{2.770262in}}%
\pgfpathlineto{\pgfqpoint{1.633255in}{2.596544in}}%
\pgfpathlineto{\pgfqpoint{1.752870in}{2.426678in}}%
\pgfpathlineto{\pgfqpoint{1.872486in}{2.260692in}}%
\pgfpathlineto{\pgfqpoint{1.986664in}{2.105915in}}%
\pgfpathlineto{\pgfqpoint{2.100842in}{1.954791in}}%
\pgfpathlineto{\pgfqpoint{2.215020in}{1.807439in}}%
\pgfpathlineto{\pgfqpoint{2.323762in}{1.670790in}}%
\pgfpathlineto{\pgfqpoint{2.427066in}{1.544572in}}%
\pgfpathlineto{\pgfqpoint{2.519496in}{1.434966in}}%
\pgfpathlineto{\pgfqpoint{2.606489in}{1.335169in}}%
\pgfpathlineto{\pgfqpoint{2.682608in}{1.251140in}}%
\pgfpathlineto{\pgfqpoint{2.747852in}{1.182275in}}%
\pgfpathlineto{\pgfqpoint{2.802223in}{1.127850in}}%
\pgfpathlineto{\pgfqpoint{2.851156in}{1.081969in}}%
\pgfpathlineto{\pgfqpoint{2.889216in}{1.048993in}}%
\pgfpathlineto{\pgfqpoint{2.921838in}{1.023184in}}%
\pgfpathlineto{\pgfqpoint{2.954460in}{1.000253in}}%
\pgfpathlineto{\pgfqpoint{2.981646in}{0.983854in}}%
\pgfpathlineto{\pgfqpoint{3.003394in}{0.972846in}}%
\pgfpathlineto{\pgfqpoint{3.025142in}{0.963982in}}%
\pgfpathlineto{\pgfqpoint{3.046890in}{0.957490in}}%
\pgfpathlineto{\pgfqpoint{3.068639in}{0.953553in}}%
\pgfpathlineto{\pgfqpoint{3.090387in}{0.952277in}}%
\pgfpathlineto{\pgfqpoint{3.112135in}{0.953674in}}%
\pgfpathlineto{\pgfqpoint{3.133883in}{0.957656in}}%
\pgfpathlineto{\pgfqpoint{3.155632in}{0.964050in}}%
\pgfpathlineto{\pgfqpoint{3.177380in}{0.972623in}}%
\pgfpathlineto{\pgfqpoint{3.204565in}{0.985998in}}%
\pgfpathlineto{\pgfqpoint{3.231750in}{1.001847in}}%
\pgfpathlineto{\pgfqpoint{3.264373in}{1.023465in}}%
\pgfpathlineto{\pgfqpoint{3.302432in}{1.051411in}}%
\pgfpathlineto{\pgfqpoint{3.351366in}{1.090385in}}%
\pgfpathlineto{\pgfqpoint{3.411173in}{1.141025in}}%
\pgfpathlineto{\pgfqpoint{3.498166in}{1.217927in}}%
\pgfpathlineto{\pgfqpoint{3.672152in}{1.375605in}}%
\pgfpathlineto{\pgfqpoint{3.867886in}{1.551902in}}%
\pgfpathlineto{\pgfqpoint{4.014687in}{1.681138in}}%
\pgfpathlineto{\pgfqpoint{4.150613in}{1.797917in}}%
\pgfpathlineto{\pgfqpoint{4.286540in}{1.911716in}}%
\pgfpathlineto{\pgfqpoint{4.417029in}{2.018091in}}%
\pgfpathlineto{\pgfqpoint{4.547519in}{2.121645in}}%
\pgfpathlineto{\pgfqpoint{4.678008in}{2.222414in}}%
\pgfpathlineto{\pgfqpoint{4.808498in}{2.320455in}}%
\pgfpathlineto{\pgfqpoint{4.944424in}{2.419765in}}%
\pgfpathlineto{\pgfqpoint{5.080351in}{2.516306in}}%
\pgfpathlineto{\pgfqpoint{5.221714in}{2.613902in}}%
\pgfpathlineto{\pgfqpoint{5.363078in}{2.708784in}}%
\pgfpathlineto{\pgfqpoint{5.509878in}{2.804617in}}%
\pgfpathlineto{\pgfqpoint{5.662116in}{2.901305in}}%
\pgfpathlineto{\pgfqpoint{5.819791in}{2.998802in}}%
\pgfpathlineto{\pgfqpoint{5.988339in}{3.100359in}}%
\pgfpathlineto{\pgfqpoint{6.167762in}{3.205842in}}%
\pgfpathlineto{\pgfqpoint{6.320000in}{3.293555in}}%
\pgfpathlineto{\pgfqpoint{6.320000in}{3.293555in}}%
\pgfusepath{stroke}%
\end{pgfscope}%
\begin{pgfscope}%
\pgfpathrectangle{\pgfqpoint{0.888378in}{0.718012in}}{\pgfqpoint{5.431622in}{3.433817in}}%
\pgfusepath{clip}%
\pgfsetbuttcap%
\pgfsetroundjoin%
\definecolor{currentfill}{rgb}{0.862745,0.078431,0.235294}%
\pgfsetfillcolor{currentfill}%
\pgfsetlinewidth{1.003750pt}%
\definecolor{currentstroke}{rgb}{0.000000,0.000000,0.000000}%
\pgfsetstrokecolor{currentstroke}%
\pgfsetdash{}{0pt}%
\pgfsys@defobject{currentmarker}{\pgfqpoint{-0.077641in}{-0.077641in}}{\pgfqpoint{0.077641in}{0.077641in}}{%
\pgfpathmoveto{\pgfqpoint{0.000000in}{-0.077641in}}%
\pgfpathcurveto{\pgfqpoint{0.020591in}{-0.077641in}}{\pgfqpoint{0.040341in}{-0.069460in}}{\pgfqpoint{0.054901in}{-0.054901in}}%
\pgfpathcurveto{\pgfqpoint{0.069460in}{-0.040341in}}{\pgfqpoint{0.077641in}{-0.020591in}}{\pgfqpoint{0.077641in}{0.000000in}}%
\pgfpathcurveto{\pgfqpoint{0.077641in}{0.020591in}}{\pgfqpoint{0.069460in}{0.040341in}}{\pgfqpoint{0.054901in}{0.054901in}}%
\pgfpathcurveto{\pgfqpoint{0.040341in}{0.069460in}}{\pgfqpoint{0.020591in}{0.077641in}}{\pgfqpoint{0.000000in}{0.077641in}}%
\pgfpathcurveto{\pgfqpoint{-0.020591in}{0.077641in}}{\pgfqpoint{-0.040341in}{0.069460in}}{\pgfqpoint{-0.054901in}{0.054901in}}%
\pgfpathcurveto{\pgfqpoint{-0.069460in}{0.040341in}}{\pgfqpoint{-0.077641in}{0.020591in}}{\pgfqpoint{-0.077641in}{0.000000in}}%
\pgfpathcurveto{\pgfqpoint{-0.077641in}{-0.020591in}}{\pgfqpoint{-0.069460in}{-0.040341in}}{\pgfqpoint{-0.054901in}{-0.054901in}}%
\pgfpathcurveto{\pgfqpoint{-0.040341in}{-0.069460in}}{\pgfqpoint{-0.020591in}{-0.077641in}}{\pgfqpoint{0.000000in}{-0.077641in}}%
\pgfpathlineto{\pgfqpoint{0.000000in}{-0.077641in}}%
\pgfpathclose%
\pgfusepath{stroke,fill}%
}%
\begin{pgfscope}%
\pgfsys@transformshift{1.341013in}{3.027732in}%
\pgfsys@useobject{currentmarker}{}%
\end{pgfscope}%
\end{pgfscope}%
\begin{pgfscope}%
\pgfpathrectangle{\pgfqpoint{0.888378in}{0.718012in}}{\pgfqpoint{5.431622in}{3.433817in}}%
\pgfusepath{clip}%
\pgfsetbuttcap%
\pgfsetroundjoin%
\definecolor{currentfill}{rgb}{0.862745,0.078431,0.235294}%
\pgfsetfillcolor{currentfill}%
\pgfsetlinewidth{1.003750pt}%
\definecolor{currentstroke}{rgb}{0.000000,0.000000,0.000000}%
\pgfsetstrokecolor{currentstroke}%
\pgfsetdash{}{0pt}%
\pgfsys@defobject{currentmarker}{\pgfqpoint{-0.099068in}{-0.084273in}}{\pgfqpoint{0.099068in}{0.104167in}}{%
\pgfpathmoveto{\pgfqpoint{0.000000in}{0.104167in}}%
\pgfpathlineto{\pgfqpoint{-0.023387in}{0.032189in}}%
\pgfpathlineto{\pgfqpoint{-0.099068in}{0.032189in}}%
\pgfpathlineto{\pgfqpoint{-0.037841in}{-0.012295in}}%
\pgfpathlineto{\pgfqpoint{-0.061228in}{-0.084273in}}%
\pgfpathlineto{\pgfqpoint{-0.000000in}{-0.039788in}}%
\pgfpathlineto{\pgfqpoint{0.061228in}{-0.084273in}}%
\pgfpathlineto{\pgfqpoint{0.037841in}{-0.012295in}}%
\pgfpathlineto{\pgfqpoint{0.099068in}{0.032189in}}%
\pgfpathlineto{\pgfqpoint{0.023387in}{0.032189in}}%
\pgfpathlineto{\pgfqpoint{0.000000in}{0.104167in}}%
\pgfpathclose%
\pgfusepath{stroke,fill}%
}%
\begin{pgfscope}%
\pgfsys@transformshift{3.089845in}{0.952276in}%
\pgfsys@useobject{currentmarker}{}%
\end{pgfscope}%
\end{pgfscope}%
\begin{pgfscope}%
\pgfsetrectcap%
\pgfsetmiterjoin%
\pgfsetlinewidth{0.803000pt}%
\definecolor{currentstroke}{rgb}{0.000000,0.000000,0.000000}%
\pgfsetstrokecolor{currentstroke}%
\pgfsetdash{}{0pt}%
\pgfpathmoveto{\pgfqpoint{0.888378in}{0.718012in}}%
\pgfpathlineto{\pgfqpoint{0.888378in}{4.151828in}}%
\pgfusepath{stroke}%
\end{pgfscope}%
\begin{pgfscope}%
\pgfsetrectcap%
\pgfsetmiterjoin%
\pgfsetlinewidth{0.803000pt}%
\definecolor{currentstroke}{rgb}{0.000000,0.000000,0.000000}%
\pgfsetstrokecolor{currentstroke}%
\pgfsetdash{}{0pt}%
\pgfpathmoveto{\pgfqpoint{6.320000in}{0.718012in}}%
\pgfpathlineto{\pgfqpoint{6.320000in}{4.151828in}}%
\pgfusepath{stroke}%
\end{pgfscope}%
\begin{pgfscope}%
\pgfsetrectcap%
\pgfsetmiterjoin%
\pgfsetlinewidth{0.803000pt}%
\definecolor{currentstroke}{rgb}{0.000000,0.000000,0.000000}%
\pgfsetstrokecolor{currentstroke}%
\pgfsetdash{}{0pt}%
\pgfpathmoveto{\pgfqpoint{0.888378in}{0.718012in}}%
\pgfpathlineto{\pgfqpoint{6.320000in}{0.718012in}}%
\pgfusepath{stroke}%
\end{pgfscope}%
\begin{pgfscope}%
\pgfsetrectcap%
\pgfsetmiterjoin%
\pgfsetlinewidth{0.803000pt}%
\definecolor{currentstroke}{rgb}{0.000000,0.000000,0.000000}%
\pgfsetstrokecolor{currentstroke}%
\pgfsetdash{}{0pt}%
\pgfpathmoveto{\pgfqpoint{0.888378in}{4.151828in}}%
\pgfpathlineto{\pgfqpoint{6.320000in}{4.151828in}}%
\pgfusepath{stroke}%
\end{pgfscope}%
\begin{pgfscope}%
\definecolor{textcolor}{rgb}{0.000000,0.000000,0.000000}%
\pgfsetstrokecolor{textcolor}%
\pgfsetfillcolor{textcolor}%
\pgftext[x=3.604189in,y=4.406852in,,base]{\color{textcolor}\sffamily\fontsize{22.000000}{26.400000}\selectfont \(\displaystyle \textrm{Polynomial manifold for 2 cycles}\)}%
\end{pgfscope}%
\begin{pgfscope}%
\pgfsetbuttcap%
\pgfsetmiterjoin%
\definecolor{currentfill}{rgb}{1.000000,1.000000,1.000000}%
\pgfsetfillcolor{currentfill}%
\pgfsetfillopacity{0.800000}%
\pgfsetlinewidth{1.003750pt}%
\definecolor{currentstroke}{rgb}{0.800000,0.800000,0.800000}%
\pgfsetstrokecolor{currentstroke}%
\pgfsetstrokeopacity{0.800000}%
\pgfsetdash{}{0pt}%
\pgfpathmoveto{\pgfqpoint{3.070788in}{3.272662in}}%
\pgfpathlineto{\pgfqpoint{6.164444in}{3.272662in}}%
\pgfpathquadraticcurveto{\pgfqpoint{6.208889in}{3.272662in}}{\pgfqpoint{6.208889in}{3.317107in}}%
\pgfpathlineto{\pgfqpoint{6.208889in}{3.996273in}}%
\pgfpathquadraticcurveto{\pgfqpoint{6.208889in}{4.040717in}}{\pgfqpoint{6.164444in}{4.040717in}}%
\pgfpathlineto{\pgfqpoint{3.070788in}{4.040717in}}%
\pgfpathquadraticcurveto{\pgfqpoint{3.026344in}{4.040717in}}{\pgfqpoint{3.026344in}{3.996273in}}%
\pgfpathlineto{\pgfqpoint{3.026344in}{3.317107in}}%
\pgfpathquadraticcurveto{\pgfqpoint{3.026344in}{3.272662in}}{\pgfqpoint{3.070788in}{3.272662in}}%
\pgfpathlineto{\pgfqpoint{3.070788in}{3.272662in}}%
\pgfpathclose%
\pgfusepath{stroke,fill}%
\end{pgfscope}%
\begin{pgfscope}%
\pgfsetbuttcap%
\pgfsetroundjoin%
\definecolor{currentfill}{rgb}{0.862745,0.078431,0.235294}%
\pgfsetfillcolor{currentfill}%
\pgfsetlinewidth{1.003750pt}%
\definecolor{currentstroke}{rgb}{0.000000,0.000000,0.000000}%
\pgfsetstrokecolor{currentstroke}%
\pgfsetdash{}{0pt}%
\pgfsys@defobject{currentmarker}{\pgfqpoint{-0.077641in}{-0.077641in}}{\pgfqpoint{0.077641in}{0.077641in}}{%
\pgfpathmoveto{\pgfqpoint{0.000000in}{-0.077641in}}%
\pgfpathcurveto{\pgfqpoint{0.020591in}{-0.077641in}}{\pgfqpoint{0.040341in}{-0.069460in}}{\pgfqpoint{0.054901in}{-0.054901in}}%
\pgfpathcurveto{\pgfqpoint{0.069460in}{-0.040341in}}{\pgfqpoint{0.077641in}{-0.020591in}}{\pgfqpoint{0.077641in}{0.000000in}}%
\pgfpathcurveto{\pgfqpoint{0.077641in}{0.020591in}}{\pgfqpoint{0.069460in}{0.040341in}}{\pgfqpoint{0.054901in}{0.054901in}}%
\pgfpathcurveto{\pgfqpoint{0.040341in}{0.069460in}}{\pgfqpoint{0.020591in}{0.077641in}}{\pgfqpoint{0.000000in}{0.077641in}}%
\pgfpathcurveto{\pgfqpoint{-0.020591in}{0.077641in}}{\pgfqpoint{-0.040341in}{0.069460in}}{\pgfqpoint{-0.054901in}{0.054901in}}%
\pgfpathcurveto{\pgfqpoint{-0.069460in}{0.040341in}}{\pgfqpoint{-0.077641in}{0.020591in}}{\pgfqpoint{-0.077641in}{0.000000in}}%
\pgfpathcurveto{\pgfqpoint{-0.077641in}{-0.020591in}}{\pgfqpoint{-0.069460in}{-0.040341in}}{\pgfqpoint{-0.054901in}{-0.054901in}}%
\pgfpathcurveto{\pgfqpoint{-0.040341in}{-0.069460in}}{\pgfqpoint{-0.020591in}{-0.077641in}}{\pgfqpoint{0.000000in}{-0.077641in}}%
\pgfpathlineto{\pgfqpoint{0.000000in}{-0.077641in}}%
\pgfpathclose%
\pgfusepath{stroke,fill}%
}%
\begin{pgfscope}%
\pgfsys@transformshift{3.337455in}{3.830378in}%
\pgfsys@useobject{currentmarker}{}%
\end{pgfscope}%
\end{pgfscope}%
\begin{pgfscope}%
\definecolor{textcolor}{rgb}{0.000000,0.000000,0.000000}%
\pgfsetstrokecolor{textcolor}%
\pgfsetfillcolor{textcolor}%
\pgftext[x=3.581899in,y=3.772044in,left,base]{\color{textcolor}\sffamily\fontsize{16.000000}{19.200000}\selectfont \(\displaystyle \textrm{Leapfrog~\cite{Verlet:1967}} \; (q=1)\)}%
\end{pgfscope}%
\begin{pgfscope}%
\pgfsetbuttcap%
\pgfsetroundjoin%
\definecolor{currentfill}{rgb}{0.862745,0.078431,0.235294}%
\pgfsetfillcolor{currentfill}%
\pgfsetlinewidth{1.003750pt}%
\definecolor{currentstroke}{rgb}{0.000000,0.000000,0.000000}%
\pgfsetstrokecolor{currentstroke}%
\pgfsetdash{}{0pt}%
\pgfsys@defobject{currentmarker}{\pgfqpoint{-0.099068in}{-0.084273in}}{\pgfqpoint{0.099068in}{0.104167in}}{%
\pgfpathmoveto{\pgfqpoint{0.000000in}{0.104167in}}%
\pgfpathlineto{\pgfqpoint{-0.023387in}{0.032189in}}%
\pgfpathlineto{\pgfqpoint{-0.099068in}{0.032189in}}%
\pgfpathlineto{\pgfqpoint{-0.037841in}{-0.012295in}}%
\pgfpathlineto{\pgfqpoint{-0.061228in}{-0.084273in}}%
\pgfpathlineto{\pgfqpoint{-0.000000in}{-0.039788in}}%
\pgfpathlineto{\pgfqpoint{0.061228in}{-0.084273in}}%
\pgfpathlineto{\pgfqpoint{0.037841in}{-0.012295in}}%
\pgfpathlineto{\pgfqpoint{0.099068in}{0.032189in}}%
\pgfpathlineto{\pgfqpoint{0.023387in}{0.032189in}}%
\pgfpathlineto{\pgfqpoint{0.000000in}{0.104167in}}%
\pgfpathclose%
\pgfusepath{stroke,fill}%
}%
\begin{pgfscope}%
\pgfsys@transformshift{3.337455in}{3.479684in}%
\pgfsys@useobject{currentmarker}{}%
\end{pgfscope}%
\end{pgfscope}%
\begin{pgfscope}%
\definecolor{textcolor}{rgb}{0.000000,0.000000,0.000000}%
\pgfsetstrokecolor{textcolor}%
\pgfsetfillcolor{textcolor}%
\pgftext[x=3.581899in,y=3.421350in,left,base]{\color{textcolor}\sffamily\fontsize{16.000000}{19.200000}\selectfont \(\displaystyle \textrm{Omelyan \textit{et al.}~\cite{OMELYAN2003272}} \; (q=2)\)}%
\end{pgfscope}%
\end{pgfpicture}%
\makeatother%
\endgroup%

%% file: novel_schemes.tex
% !TEX root = Efficient_Trotterizations

\section{Novel scheme results}\label{sec:novel_schemes}

In this section we discuss the results of our scheme optimization simulations and their performance in numerical experiments of the Heisenberg XXZ model and the quantum harmonic oscillator.
Importantly, we only consider real-valued schemes $a_{i}, b_{i} \in \mathds{R}$, due to their unitarity conservation.
However, the framework allows for optimization over the complex plane, and much could be discovered in this scheme parameter region.
With that said, using the developed framework we were able to (hopefully) fully explore real-valued schemes of orders $n = 2, 4, 6$ for each number of cycles $q$.
At each $q$ we found many minima of the theoretical error function $\textrm{Err}_{n}$ as in equation~\eqref{eq:order_error}, and identified the global minimum.
Running the minimizer repeatedly in the relevant region, we consistently found the same global minima.
With each run we also found fewer new local minima so that we are confident to have identified all the relevant solutions.
In order to accurately compare schemes of a certain order, we need to properly define the efficiency $\textrm{Eff}_{n}$.
Following Ref.~\cite{OMELYAN2003272}, we can define it by how small the leading errors of a scheme $\textrm{Err}_{n}$~\eqref{eq:order_error} are compared to the number of cycles $q$ they require,

\begin{equation}
	\textrm{Eff}_{n} = \frac{1}{q^{n} \textrm{Err}_{n}}.\label{eq:eff}
\end{equation}

\noindent
More specifically, the global error scales as $ h^{n}\textrm{Err}_{n}$ and for a fixed computational cost the step size needs to be scaled with the number of cycles $h\propto q$.
Thus, the efficiency in the definition~\eqref{eq:eff} encodes the cost-independent scheme-specific proportionality factor of the error $\propto  q^n\textrm{Err}_{n}$.
%The number of cycles in the definition~\eqref{eq:eff} scale with the order $n$ because the global error scales as $\mathcal{O} (h^{n})$, which leaves the efficiency $\textrm{Eff}_{n}$ constant.
In the following, we first present the results of the theoretical efficiency, and give our recommendations for the best-performing schemes.

\begin{figure}[h!]
	\begin{center}
    \resizebox{0.95\textwidth}{!}{%
		\input{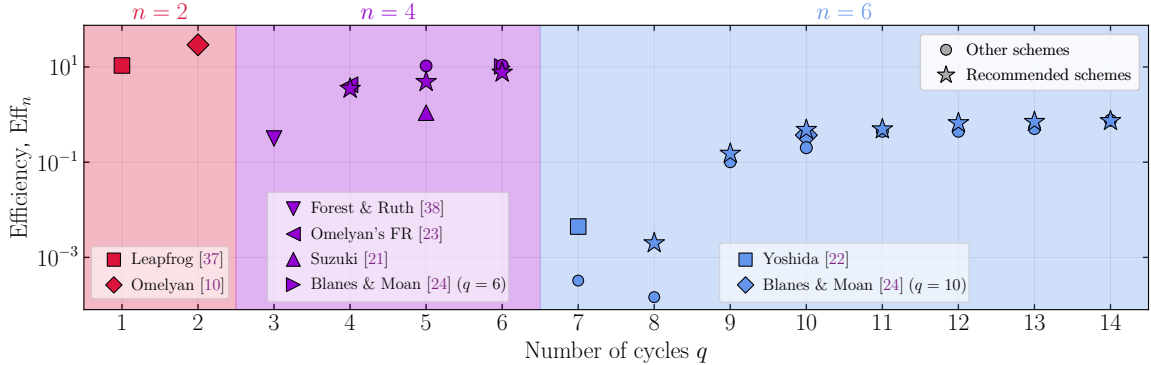}
	  }
	\end{center}
	\caption{Theoretical efficiency $\textrm{Eff}_{n}$ according to Eq.~\eqref{eq:eff} at orders $n = 2, 4, 6$ across the relevant number of cycles $q$ for a collection of historical schemes and our novel decompositions.
	         We find a plateau towards the maximal number of cycles in a given order.
           For our novel schemes we plot a recommended one according to its theoretical efficiency and consistent performance in practice.
           In order to compare our best schemes, we also plot the scheme with the highest $\textrm{Eff}_{n}$, unless that is our recommended one in which case we plot the theoretically next best decomposition.
           From the results at the maximal no.\ cycles we find a slight theoretical improvement of our schemes in both orders $n = 4$ and $n = 6$ over those by Blanes \& Moan~\cite{BLANES2002313}.
		       We note that the efficiency values are not comparable between orders.}\label{fig:theo_efficiency}
\end{figure}

The framework was first benchmarked by recovering the historical schemes found by our predecessors.
Afterwards, it was run in the unexplored region of cycles $q$, and we present the theoretical efficiency of these decompositions along with our novel schemes across orders $n = 2, 4, 6$ in \Cref{fig:theo_efficiency}.
We plot them according to their efficiency $\textrm{Eff}_{n}$ at order $n$, where we note that the scale of efficiency values is not comparable between orders.
For each number of cycles $q$ we recommend a scheme according to their theoretical efficiency and practical performance.
We also plot the Trotterization with the highest efficiency, unless that is our recommended scheme, in which case this point is the next best one.
This is done in order to compare the recommended schemes to other decompositions.
We find an improvement over the previously known Trotterization and that it is possible to maximally improve on the theoretical efficiency by going to the highest number of cycles $q$ for orders $n \leq 6$.

\newpage

Thus, we recommend the use of our schemes at $q = 6$ for order $n = 4$ and $q = 14$ for order $n = 6$ Trotterized time evolution.
We provide the parameters $c_{i}$ and $d_{i}$ for these in Tables~\ref{tab:recommened6} and~\ref{tab:recommened14} respectively.
Although, we only present the parameters for these two schemes here, the other decompositions can be found on our repository~\cite{markomalezic_2026_18347430}.
The two recommended schemes closely follow the theoretically most efficient decompositions, which turn out to perform worse in practice.
We speculate this is due to the distance from the origin point of the scheme parameters, while the recommended schemes live relatively close to this point.
The origin point is defined at $c_{i} = d_{i} = \nicefrac{1}{2 q}$, and the distance from it as the following norm,

\begin{equation}
  \bar{x}^{2} = 2 \sum_{i}^{q} \left( c_{i} - \frac{1}{2 q} \right)^{2},\label{eq:origin}
\end{equation}

\noindent
where we combined the parameters $c_{i}$ and $d_{i}$ due to their symmetric properties.
We present the distance from the origin point for our recommended schemes in \Cref{tab:scheme_comparison}, and compare them to the theoretically most efficient schemes and the ones by Blanes $\&$ Moan.
While the improvement is marginal at $4^{\textrm{th}}$ order, it is significant at $6^{\textrm{th}}$ order, which translates into better practical performance.
The proximity to the origin is studied further in \Cref{sec:uniformity} in hopes of improving practical efficiency.
The recommendation of these schemes is thus justified by these theoretical reasons and their practical performance, which we discuss in the following.

\begin{table}[h!]
	\centering
	\begin{minipage}[t]{0.48\textwidth}
		\centering
		\captionof{table}{Recommended parameters $c_{i}=d_{q+1-i}$ for an order $n = 4$ scheme at $q = 6$ cycles.
                      The scheme has the second-highest efficiency $\textrm{Eff}_{4}$ at this order, and performs well due to its proximity to the origin point.}\label{tab:recommened6}
		\vspace{2.0cm}
		\begin{tabular}{C{0.5cm}C{5.75cm}}
			\toprule
			$i$ &                   $c_{i} = d_{q+1-i}$   \\
			\midrule
			1  &                  $0.074082572180463262$  \\
			2  &                  $0.232923088374338803$  \\
			3  &                  $0.296820560634668408$  \\
			4  &                  $0.122086989386933251$  \\
			5  & \hspace{-0.45cm} $-0.350153632343424469$ \\
			6  &                  $0.124240421767020743$  \\
			\bottomrule
		\end{tabular}
	\end{minipage}
	\hfill
	\begin{minipage}[t]{0.48\textwidth}
		\centering
		\captionof{table}{Recommended parameters $c_{i}=d_{q+1-i}$ for an order $n = 6$ scheme at $q = 14$ cycles.
                      The scheme has the second-highest efficiency $\textrm{Eff}_{6}$ at this order, and performs well due to its proximity to the origin point.}\label{tab:recommened14}
		\begin{tabular}{C{0.5cm}C{5.75cm}}
			\toprule
			$i$ &                   $c_{i} = d_{q+1-i}$   \\
			\midrule
			1  &                  $0.037251326545569924$  \\
			2  &                  $0.120600278793781562$  \\
			3  &                  $0.266062994460763541$  \\
			4  &                  $0.163668553338143183$  \\
			5  &                  $0.071316838327437583$  \\
			6  &                  $0.058117508592333414$  \\
			7  &                  $0.188707697234255120$  \\
			8  & \hspace{-0.45cm} $-0.200016005078878524$ \\
			9  &                  $0.074145714537530386$  \\
			10 &                  $0.087345801243357893$  \\
			11 &                  $0.044234977360777830$  \\
			12 & \hspace{-0.45cm} $-0.230821838291030424$  \\
			13 & \hspace{-0.45cm} $-0.237197828922049295$  \\
			14 &                  $0.056583981858007803$  \\
			\bottomrule
		\end{tabular}
	\end{minipage}
\end{table}

%\begin{table}[h!]
%  \centering
%  \caption{Comparison of our recommended schemes, the theoretically most efficient schemes and schemes by Blanes \& Moan at $4^{\textrm{th}}$ and $6^{\textrm{th}}$ order.
%           Higher theoretical efficiency $\textrm{Eff}_{n}$ and a smaller distance from the origin $\bar{x}$ are two metrics, which are believed to contribute most to the practical performance of product formulas.
%           At order $n = 4$, the differences between the metrics are marginal while at order $n = 6$ the improvement of our new-found schemes is significant.
%           The efficiency of the top most schemes is roughly twice as high as the one recommended by Blanes $\&$ Moan, and the distance from the origin is about twice as small for the scheme we recommend.
%           This is also the reason for its recommendation as opposed to the one with the highest efficiency.}\label{tab:scheme_comparison}
%  \setlength{\tabcolsep}{12pt}
%  \begin{tabular}{c c S S S}
%    \toprule
%    $n$ & Metric & {Blanes $\&$ Moan} & {Highest $\textrm{Eff}_{n}$} & {Recommended} \\
%    \midrule
%    \multirow{2}{*}{4} & $\text{Eff}_4$ & 10.216 & 10.831 & 10.530 \\
%    & $\bar{x}$ & 0.753 & 0.863 & 0.720 \\
%    \midrule
%    \multirow{2}{*}{6} & $\text{Eff}_6$ & 0.370 & 0.762 & 0.736 \\
%    & $\bar{x}$ & 1.453 & 1.256 & 0.784 \\
%    \bottomrule
%  \end{tabular}
%\end{table}

\begin{table}[h!]
  \centering
  \caption{Comparison of our recommended schemes, the theoretically most efficient schemes and schemes by Blanes \& Moan at $4^{\textrm{th}}$ and $6^{\textrm{th}}$ order.
           Higher theoretical efficiency $\textrm{Eff}_{n}$ and a smaller distance from the origin $\bar{x}$ are two metrics, which are believed to contribute most to the practical performance of product formulas.
           At order $n = 4$, the differences between the metrics are marginal while at order $n = 6$ the improvement of our new-found schemes is significant.
           The efficiency of the top most schemes is roughly twice as high as the one recommended by Blanes $\&$ Moan, and the distance from the origin is about twice as small for the scheme we recommend.
           This is also the reason for its recommendation as opposed to the one with the highest efficiency.}\label{tab:scheme_comparison}
  \setlength{\tabcolsep}{12pt}
  \begin{tabular}{c c r r r}
    \toprule
    $n$ & Metric & {Blanes $\&$ Moan} & {Highest $\textrm{Eff}_{n}$} & {Recommended} \\
    \midrule
    \multirow{2}{*}{4} & $\text{Eff}_4$ & \hspace{-1.0cm} 10.216 & 10.831 & 10.530 \\
    & $\bar{x}$ & \hspace{-1.0cm} 0.753 & 0.863 & 0.720 \\
    \midrule
    \multirow{2}{*}{6} & $\text{Eff}_6$ & 0.370 & 0.762 & 0.736 \\
    & $\bar{x}$ & 1.453 & 1.256 & 0.784 \\
    \bottomrule
  \end{tabular}
\end{table}

\newpage

\subsection{Numerical experiments -- The Heisenberg model}\label{sec:novel_schemes:Heisenberg}

\indent
We now focus on the results from our numerical experiments with the new-found decompositions.
First we studied how our schemes performed on the Heisenberg model, defined by the following Hamiltonian over a periodic spin chain of length $L$:

\begin{equation}
	H = \sum_{i = 1}^{L} \left( J^{x} \sigma_{i}^{x} \sigma_{i+1}^{x} + J^{y} \sigma_{i}^{y} \sigma_{i+1}^{y} + J^{z} \sigma_{i}^{z} \sigma_{i+1}^{z} + h_{i} \sigma_{i}^{z} \right),
\end{equation}

\noindent
where $J^{\alpha}$ represent the couplings and $\sigma^{\alpha}_{i}$ the Pauli matrices at spin site $i$.
We chose such a Hamiltonian due to its application on quantum computers, because Pauli matrices can easily be implemented as local gates.
For the following results we choose the couplings as $J^{x} = J^{y} = J^{z} = 1$ and sample the magnetic field $h_{i} \in [-0.1, 0.1]$ from a uniform distribution, which defines the Heisenberg XXZ model.
However, we also performed experiments with $J^{y} = 0$, referred to as the Heisenberg XZ model, and found similar results.
For now we choose the length of the chain as $L = 6$, which turns out to be sufficiently large.
We can exactly diagonalize this Hamiltonian for short enough chain lengths $L$ and obtain the time evolution operator $U (-it) = \exp(-i H t)$.
By comparing the exact $U(t)$ to our Trotter-Suzuki schemes $S_{n}(h)^{t/h}$, we can estimate the accumulated Trotter error by using the Frobenius norm,

\begin{equation}
	\Delta_{n}^{\textrm{exp}} = \frac{1}{\sqrt{N}} \left\| U (t) - S_{n}(h)^{t/h} \right\|_{F} = \frac{1}{\sqrt{N}} \sqrt{\sum_{v} \left| U(t) \cdot v - S_{n}(h)^{t/h} \cdot v \right|^{2}}, \label{eq:Frobenius}
\end{equation}

\noindent
where the sum runs over the basis states $v$ of the corresponding $N$-dimensional vector space with $N=2^L$ and the $| \cdot |$ represents the Euclidean norm.

Instead of the Frobenius norm other norms like the spectral norm could be used, or as advocated in Ref.~\cite{morales2022greatly} the error of the eigenvalues.
Typically, the performance ranking of different schemes will be very similar for different norms.
We settled with the Frobenius norm here because it is computationally cheap and allows to capture the full error of the time evolution operator.
Other norms might be more suitable for quantifying the error of the time evolution of specific states.
This operator-based approach is also best compatible with our theoretical framework detailed in \Cref{sec:framework}.

Finally, we have to choose the ordering of our splitting, and depending on the simulation we investigate two different choices.
Either choice relies on local operators and can be translated directly into gates for quantum computation.
First, we group the operators by sites in the following order,

\begin{equation}
	S^{(3 L)} (ih) = e^{ih H_{1}^{x} c_{1}} e^{ih H_{1}^{y} c_{1}} e^{ih H_{1}^{z} c_{1}} e^{ih H_{2}^{x} c_{1}} e^{ih H_{2}^{y} c_{1}} e^{ih H_{2}^{z} c_{1}} \cdots e^{ih H_{1}^{z} d_{q}} e^{ih H_{1}^{y} d_{q}} e^{ih H_{1}^{x} d_{q}},
\end{equation}

\noindent
where we defined the local operator $H_{i}^{\alpha} = J^{\alpha} \sigma_{i}^{\alpha} \sigma_{i+1}^{\alpha} + \delta_{a z} h_{i} \sigma_{i}^{z}$.
By counting the number of operators needed for a single ramp, we see that we need $3 L$ such stages.
However, we can reduce the number of stages by using a global grouping of similar stages in the following way,

\begin{equation}
  \begin{aligned}
    S^{(3)} (ih) &= \left( \prod_{i=1}^{L} e^{ih H_{i}^{x} c_{1}} \right) \left( \prod_{i=1}^{L} e^{ih H_{i}^{y} c_{1}} \right) \left( \prod_{i=1}^{L} e^{ih H_{i}^{z} c_{1}} \right) \cdots \\
    &\times \left( \prod_{i=L}^{1} e^{ih H_{i}^{z} d_{q}} \right) \left( \prod_{i=L}^{1} e^{ih H_{i}^{y} d_{q}} \right) \left( \prod_{i=L}^{1} e^{ih H_{i}^{x} d_{q}} \right).
  \end{aligned}
\end{equation}

\noindent
Since the exponentials of similar type commute, we have reduced the number of stages to $3$ and obtained the $S^{(3)}$ decomposition.

The second grouping is typically more efficient in classical computations.
Generally, the efficiency of a simulation depends on the grouping and the optimal choice might vary with model and geometry of the underlying system~\cite{schubert2023trotter}.
In this work, we do not aim to explore the optimal grouping.
Rather, we deliberately decouple the choice of the grouping from the choice of the Trotter scheme.
The different groupings investigated here are used exclusively to verify whether the optimal Trotterization is independent of the grouping.

We set the evolution time to $t= h N_{t} =10$ and simulate the data at a fixed computation cost $q N_{t} = 1000$.
From the error~\eqref{eq:Frobenius} we compute the experimental efficiency, which is defined in the following way,

\begin{equation}
  \textrm{Eff}_{n}^{\textrm{exp}} = t \frac{\Omega^{n}}{\Delta_n^{\textrm{exp}}}, \quad \Omega = \Omega_{0} \frac{q}{h}.\label{eq:exp_eff}
\end{equation}

\begin{figure}[h!]
	\begin{center}
    \resizebox{0.95\textwidth}{!}{%
		\input{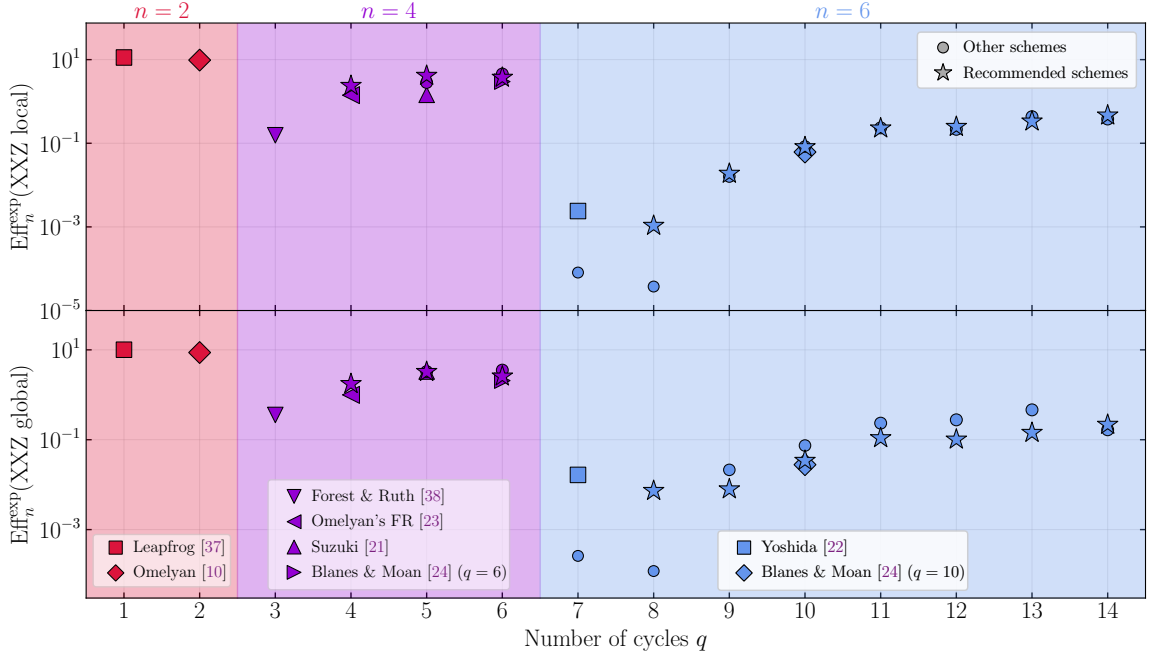}
	  }
	\end{center}
	\caption{Experimental efficiency $\textrm{Eff}_{n}^{\textrm{exp}}$~\eqref{eq:exp_eff} of the Heisenberg XXZ model with $L = 6$ spins at orders $n = 2, 4, 6$ across the relevant number of cycles $q$ for a collection of historical schemes and our novel schemes.
		We also investigated two fundamentally different scheme orderings -- local (top) and global (bottom), and found a similar picture to that of the theoretical efficiency (see Fig.~\ref{fig:theo_efficiency}).
    For our novel schemes we plot a recommended one according to its theoretical efficiency and consistent performance in practice, along with the best scheme.
    In the local grouping we find a few schemes which turn out to perform better in some cases, but are expected to perform inconsistently in other models.
    The data was simulated for total evolution time $t = 10$ at a fixed computational cost $q N_{t} = 1000$.
    The experimental efficiency is rescaled, such that it mimics the theoretical efficiency.
		We note that the efficiency values are not comparable between orders and models.}\label{fig:efficiency_Heisenberg}
\end{figure}

\noindent
It differs from the theoretical efficiency definition~\eqref{eq:eff}, because it is defined at a fixed cost~$q N_{t}$, which needs to be taken into account.
Due to the arbitrary choice of simulation cost $q N_{t}$, we add a rescaling factor $\Omega_{0}$, which we tune such that the experimental efficiency mimics the theoretical value.
We now plot the practical efficiency of the XXZ model in \Cref{fig:efficiency_Heisenberg} for the local grouping $S^{(3 L)}$ (top) and the global grouping $S^{(3)}$ (bottom).
In both cases we observe behaviour similar to the theoretical efficiency.
In practice the maximal number of cycles still proves more efficient, and our recommended order $n = 6$ scheme at $q = 14$ cycles shows an improvement over the best previously known order $n = 6$ scheme by Blanes \& Moan~\cite{BLANES2002313} in both groupings.
Though the recommended schemes perform best in the global grouping, we find that in the local grouping there exist better ones at order $n = 6$, excluding $q = 14$.
However, these schemes have a poorer theoretical efficiency, and are expected not to perform well consistently in other models.
We also observed this when performing similar simulations with the XZ model, where we found inconsistent practical efficiency of these schemes.
The only consistent result that performed well at order $n = 6$ was the recommended scheme at $q = 14$ cycles (see \Cref{tab:recommened14}).

\begin{figure}[h!]
	\centering
  \resizebox{0.99\textwidth}{!}{%
		\input{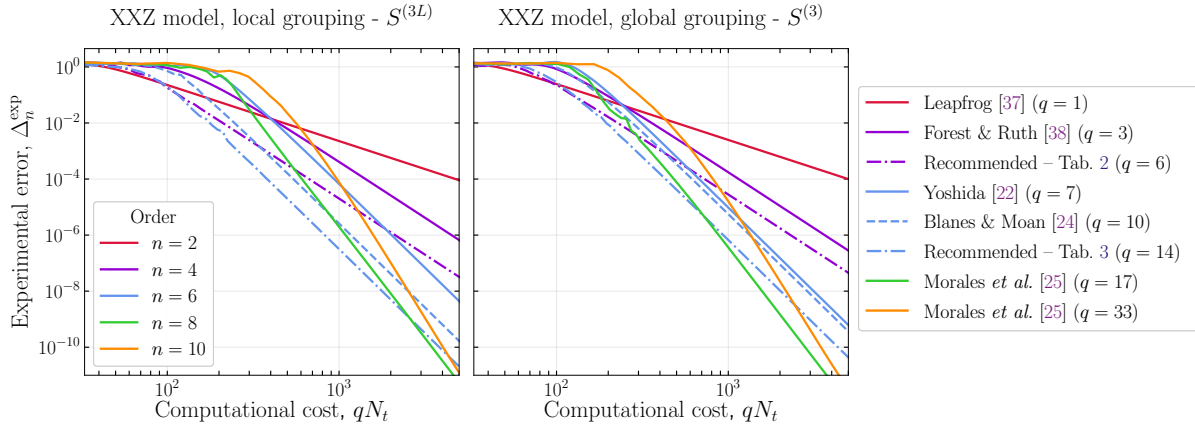}
	}
	\caption{Experimental error $\Delta_{n}^{\textrm{exp}}$ from numerical simulations of the Heisenberg XXZ model with $L = 6$ spins in the global grouping $S^{(3)}$ (left) and the local grouping $S^{(3 L)}$ (right) as a function of the computational cost $q N_{t}$.
		The data was simulated at a fixed evolution time $t = 10$.
		We plot a collection of historically relevant schemes and compare them to our recommended order $n = 4$ and $n = 6$ schemes.
    On the lower end of the cost we find a plateau, which appears due to the properties of the Frobenius norm that defines the accumulated error.
		The relevant information is in the scaling region that comes after the plateau, which increases in steepness with higher orders $n$.
		Here we find that our recommended order $n = 6$ scheme at $q = 14$ cycles performs better than the rest of the known schemes at each cost for orders $n = 2, 4, 6$ and a significant part of the cost for the best known order $n = 8, 10$ schemes by Morales \textit{et al}.~\cite{morales2022greatly}.}\label{fig:err_compXXZ}
\end{figure}

Having compared the schemes at each cycle of order $n = 2, 4, 6$, we now take our recommended schemes at $q = 6$ and $q = 14$, and compare how they fare against the historical schemes across the computational cost $q N_{t}$ for the fixed final time $t=10$.
The computational cost scales with the number of cycles $q$ due to the \textit{first same as last} property of symmetric schemes.
It also scales with the number of time steps $N_{t}$, because that is how many times a scheme $S_{n}(h)$ needs to be applied to a state in order to evolve it to some time $t$.
Plotting the experimental error for both of the groupings in \Cref{fig:err_compXXZ}, we find some interesting results.
There is a plateau on the lower end of the computational cost, where we approach the theoretical limit of the Frobenius norm as we had defined it~\eqref{eq:Frobenius}.
Afterwards, we find the scaling region, which follows the scaling law $\mathcal{O} (h^{n})=\mathcal{O} (N_t^{-n})$.
Here we find that the higher the order, the steeper the scaling, which is to be expected.
Possibly the most important result is that our recommended $q = 14$ scheme performs better than the historical schemes for orders $n \leq 6$ at each cost.
This means that even at lower costs where the Leapfrog scheme was thought to excel, we find an improvement in both of the groupings.
Furthermore, there is a significant region in the cost, where the best order $n = 8$ and $n = 10$ schemes by Morales \textit{et al}.~\cite{morales2022greatly} perform worse than our recommended order $n = 6$ scheme, though this region is somewhat reduced in the local grouping.
Since Morales \textit{et al}. employed Yoshida's method to find their schemes at minimal cycles $q$ needed to satisfy the high orders, we believe our framework could be used to significantly improve on the efficiency at higher orders as well.

\begin{figure}[h!]
	\centering
  \resizebox{0.8\textwidth}{!}{%
		\input{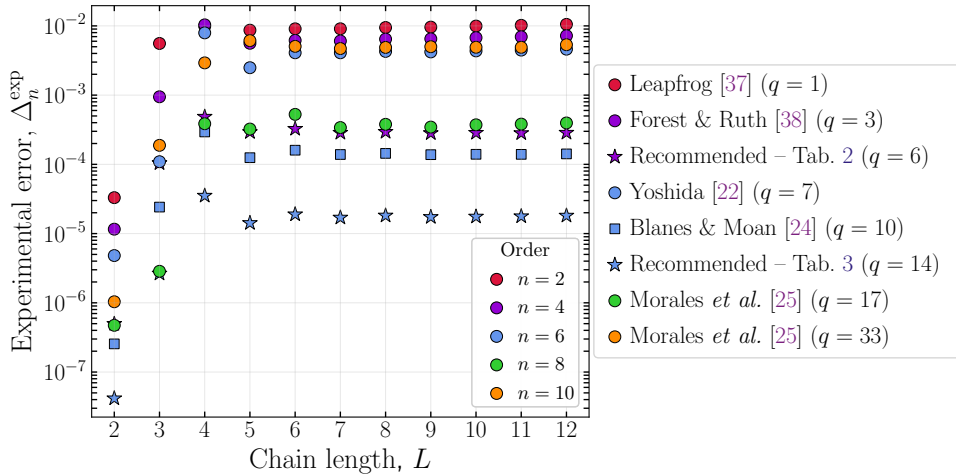}
	}
	\caption{Experimental error $\Delta_{n}^{\textrm{exp}}$ from numerical simulations of the Heisenberg XXZ model at different chain lengths $L$ in the global grouping $S^{(3)}$.
		Going towards the thermodynamic limit, we get a constant value for our collection of schemes at different orders $n$, though short chains seem less stable.
    The collection includes historically relevant schemes and our recommended schemes at orders $n = 4, 6$.
		The data was generated with a total evolution time $t = 10.0$ and computational cost $q N_{t} = 500$, which is in the scaling region for the simulated schemes (see Fig.~\ref{fig:err_compXXZ}).}\label{fig:chain}
\end{figure}

So far, we have been studying a spin chain of length $L = 6$.
It is expected for size-dependent fluctuations to average out and the relative error $\Delta_{n}^\text{exp}$ to approach a constant as $L \rightarrow \infty$.
Now it remains to verify that $L=6$ is indeed large enough to be representative for the thermodynamic limit.
To answer this, we investigated the Trotter error as one increases the length of the spin chain $L$ for the collection of schemes we looked at before (see \Cref{fig:chain}).
We chose $q N_{t} = 500$ as the computation cost, which is in the scaling region for all relevant schemes, and simulate the XXZ model in the global grouping $S^{(3)}$.
As one might have expected, short chains exhibit unpredictable fluctuations.
However, these plateau around $L \approx 5$ in the Heisenberg model, and we observe a constant value for the error across all schemes, which is expected to extend into the thermodynamic limit.
More generally, the plateau for sufficiently large $L$ is a universal feature since the novel schemes were constructed in a model-agnostic way.

An immediate implication of the findings in \Cref{fig:chain} is that the optimal Trotterization is size-independent for a given model, target time and relative error.
This opens the path to model-dependent tuning as follows.
Choose a small system with a known exact solution, for instance $L=6$ as in \Cref{fig:err_compXXZ}.
Test the performance of multiple Trotter schemes for this particular system and choose the optimal one.
Optionally verify that this scheme is still optimal for a slightly larger system.
The identified scheme can safely be used for larger systems where no reference results are available.

\newpage

\subsection{Numerical experiments -- Quantum harmonic oscillator}\label{sec:novel_schemes:Harmonic}

\begin{figure}[h!]
	\begin{center}
    \resizebox{0.95\textwidth}{!}{%
		\input{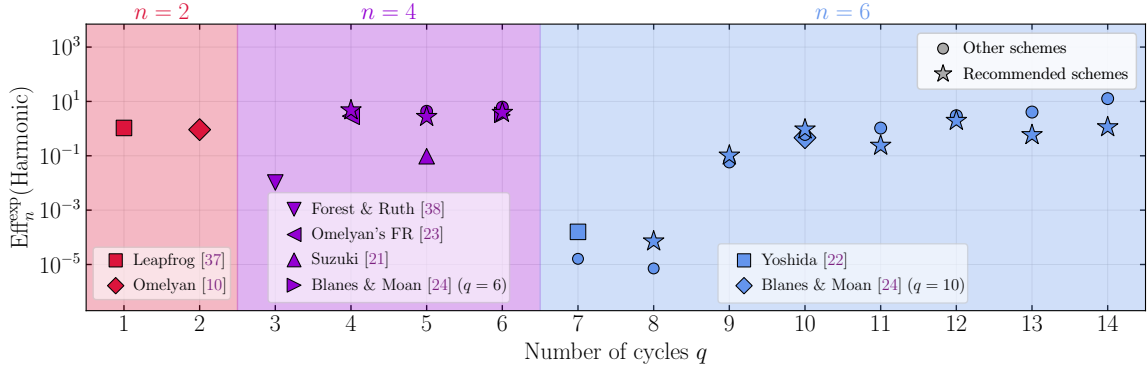}
	  }
	\end{center}
	\caption{Experimental efficiency $\textrm{Eff}_{n}^{\textrm{exp}}$~\eqref{eq:exp_eff} of the quantum harmonic oscillator at orders $n = 2, 4, 6$ across the relevant number of cycles $q$ for a collection of historical schemes and our novel schemes.
    Though our recommended schemes perform well, there exist theoretically less efficient schemes, which perform surprisingly well.
    With this observation we find a strong model dependence on the practical performance of Trotter-Suzuki schemes.
    For our novel schemes we plot a recommended one according to its theoretical efficiency and consistent performance in practice, along with the best scheme.
    The data was simulated for total evolution time $t = 200$ at a fixed computational cost $q N_{t} = 6000$.
    The experimental efficiency is rescaled, such that it mimics the theoretical efficiency.
		We note that the efficiency values are not comparable between orders and models.}\label{fig:efficiency_harmonic}
\end{figure}

In order to investigate the dependence of our schemes on the model, we also simulate the quantum harmonic oscillator, which is a symplectic system unlike the Heisenberg model~\cite{PhysRevA.37.3028}.
Also unlike the Heisenberg model, it is defined on a one dimensional space $x \in \left[-x_{0}, x_{0} \right]$, which needs to be discretized to $N_{x}$ points.
In our simulations we used $x_{0} = 10.0$ and $N_{x} = 2000$.
The Hamiltonian is the following sum of the kinetic and potential parts,

\begin{eqnarray}
	H = \frac{p^{2}}{2 m} + \frac{m}{2} \omega^{2} x^{2},
\end{eqnarray}

\noindent
where we set the parameters $m = \omega = 1.0$ and $p$ represents the momentum space.
We can estimate the Trotter error similarly to Eq.~\eqref{eq:Frobenius}, however the Hilbert space is infinitely dimensional, and we need to truncate the sum at some number of states $N_{\phi}$.
The exact time evolution of such a system is possible, because we know the eigenstates $\phi_{n}$ and energies $E_{n}$, and a time evolved state is given by,

\begin{equation}
	\psi(x, t) = \sum_{n = 0}^{N_{\textrm{cut}}} \langle \phi_{n} | \psi (x, 0) \rangle e^{-i E_{n} t} \phi_{n} (x),
\end{equation}

\noindent
where we had to again introduce a cutoff $N_{\textrm{cut}}$.
Throughout our simulations we chose $N_{\textrm{cut}} = 23$ and $N_{\phi} = 5$, because we did not find qualitative differences at higher cutoffs.
In this case we set the evolution time $t = 200$, because the short time behaviour showed significant oscillations in the accumulated error.
Setting the computational cost $q N_{t} = 6000$ and computing the efficiency in a similar manner, we find a different picture (see \Cref{fig:efficiency_harmonic}).
The schemes we recommend according to theory and the practical performance in the Heisenberg model are not always maximally efficient.
Similarly to the local grouping in the XXZ model we find that some schemes with a lower theoretical efficiency perform surprisingly well in this model.
Evidently there is a strong model dependence to the efficiency of Trotter-Suzuki schemes and this should be studied further.
Nonetheless, our recommended schemes perform sufficiently well, because of their theoretical efficiency and their proximity to the origin point.
Taking this into account, we anticipate them to perform well in general.

\begin{figure}[h!]
	\centering
  \resizebox{0.75\textwidth}{!}{%
		\input{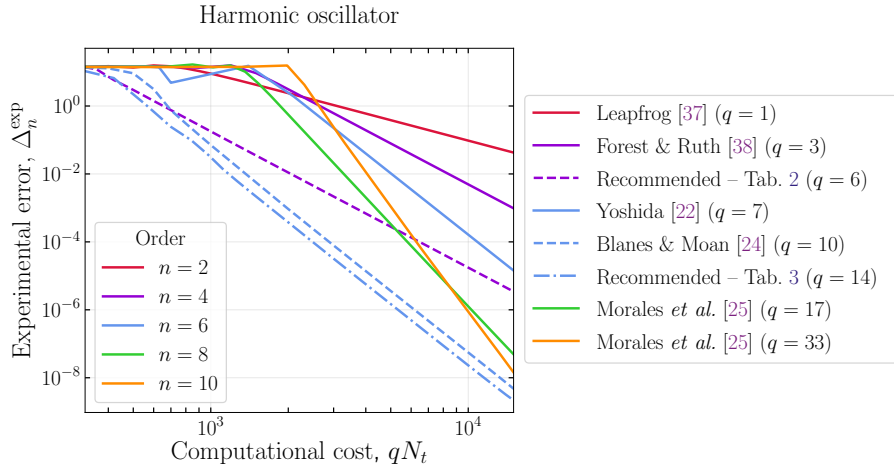}
	}
	\caption{Experimental error $\Delta_{n}^{\textrm{exp}}$ from numerical simulations of the quantum harmonic oscillator as a function of the computational cost $q N_{t}$.
		The data was simulated at a fixed evolution time $t = 200$.
		We plot a collection of historically relevant schemes and compare them to our recommended order $n = 4$ and $n = 6$ schemes.
    We find that our recommended order $n = 6$ scheme at $q = 14$ cycles performs better in the scaling region than the previously known schemes of the same or lower order for each cost $q N_{t}$.
    In a significant part of the computational cost it also performs better compared to the schemes by Morales \textit{et al.}~\cite{morales2022greatly}.
    Though we had to truncate the Frobenius norm, we still find a plateau on the lower end of the computational cost, however it does not converge to 1, as we expect in the full definition.}\label{fig:err_compHarm}
\end{figure}

In order to complete the picture, we again compare the historically relevant schemes and our recommended ones for each computational cost $q N_{t}$.
Plotting the accumulated error of the total evolution time $t = 200$ on \Cref{fig:err_compHarm}, we find similar results to that of the Heisenberg model (see Fig.~\ref{fig:err_compXXZ}).
After a plateau due to the properties of the Frobenius norm we find the scaling region, where each order scales as expected.
The recommended scheme at $q = 14$ cycles performs slightly better than the one by Blanes \& Moan~\cite{BLANES2002313}, and much better than the higher order schemes by Morales \textit{et al.}~\cite{morales2022greatly} in a wide region of the cost $q N_{t}$.

%% file: uniformity.tex
% !TEX root = Efficient_Trotterizations

\section{Practical framework improvement}\label{sec:uniformity}

Finally, we present an observation we made when analysing the found schemes.
This has to do with the fact that global minima of the error function $\textrm{Err}_{n}$, as we have defined them in Eq.~\eqref{eq:order_error}, do not necessarily perform well in practice.
It turns out that the values of the scheme parameters $c_{i}$, $d_{i}$ themselves also influence the accumulated error.
In the ideal case they would be equal, which constrains them to $c_{i} = d_{i} = \frac{1}{2 q}$, which we call the origin point.
In reality this is only achieved at $q = 1$ for the Leapfrog scheme, otherwise we do not construct a valid scheme of higher order.
Large deviations from this point not only perform worse in practice, but are usually also less efficient according to~\eqref{eq:eff}.

We give the theoretical reason for this in \Cref{sec:error_accumulation}, which is based on the work by Chen~\cite{chen2024trottererrortimescaling}.
In short, the practical performance is dominated by the error accumulation as opposed to the pure error at each small time step.
In Refs.~\cite{chen2024trottererrortimescaling,chen2024errorinterferencequantumsimulation} it has been shown that the error at every step can be split into a `parallel' and an `orthogonal' component.
The theoretical error $\textrm{Err}_n$ captures their sum of squares.
Over many steps, the orthogonal error component rotates around and averages out.
The parallel error component, on the other hand, accumulates linearly.
We are interested in Trotter schemes with favourable long time error accumulation and thus need to minimise the parallel error component.
For lower orders we can show that the parallel error component is minimized when all the coefficients $c_i,d_i$ are constant.
The same property has not been proven for higher orders, but there is clear numerical evidence~\cite{Ostmeyer:2022} we find the variance of $c_i,d_i$ to be a good indicator for the expected error accumulation.

\begin{figure}[h!]
	\begin{center}
		\resizebox{0.55\textwidth}{!}{%
		\input{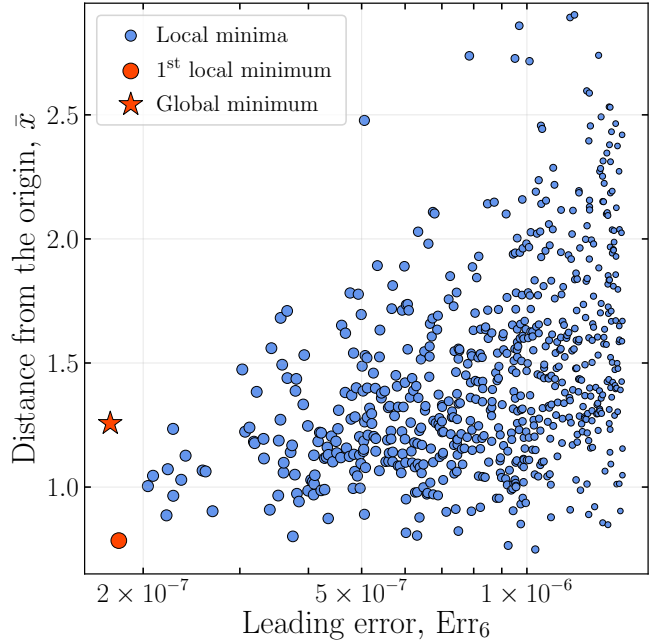}
	  }
	\end{center}
	\caption{A collection of order $n = 6$ schemes at $q = 14$ cycles plotted for their distance from the origin $\bar{x}$~\eqref{eq:origin} against the leading-order error $\textrm{Err}_{6}$.
		       We find hundreds of minima in the $\textrm{Err}_{6}$ function, but only examine them above a certain efficiency threshold, which is why there exist a cutoff on the higher end of $\textrm{Err}_{6}$.
           We also find a cutoff on the lower end of the distance $\bar{x}_{\textrm{min}}$, which appears naturally due to additional parameter constraints at higher orders.
           The global and $1^{\textrm{st}}$ local minima are highlighted because of their similar leading-order error, but difference in the distance $\bar{x}$.
           The proximity of the $1^{\textrm{st}}$ local minimum to the origin is the reason for its consistent performance, and why we recommend it (see Tab.~\ref{tab:recommened14}).}\label{fig:origin_err}
\end{figure}

Here we present the observation on the order $n = 6$ schemes at $q = 14$ cycles, for which we compute the leading-order error $\textrm{Err}_{6}$ similarly to~\eqref{eq:order_error}.
We compare this to the distance from the origin $\bar{x}$ as defined in Eq.~\eqref{eq:origin} of each scheme in order to understand how the two quantities impact the practical performance.
In \Cref{fig:origin_err} we plot these two quantities against each other for all minima found in the optimization.
Because we only examined schemes above some efficiency $\textrm{Eff}_{6}$, we produce a cutoff on the higher end of $\textrm{Err}_{6}$.
There also appears to be a minimal $\bar{x}_{\textrm{min}}$, which is believed to exist due to the properties of higher order schemes, constraining a certain number of parameters $c_{i}$ and $d_{i}$ to negative values.
We highlight the $\textrm{Err}_{6}$ global minimum and the $1^{\textrm{st}}$ local minimum, which is also our recommended $6^{\textrm{th}}$ order scheme (see \Cref{tab:recommened14}).
The two schemes have a similar leading-order error $\textrm{Err}_{6}$, but different distance from the origin $\bar{x}$.
The $1^{\textrm{st}}$ local minimum is closer to the origin than the global minimum, and thus performs better in practice.

\begin{figure}[h!]
	\begin{center}
		\resizebox{0.99\textwidth}{!}{%
		\input{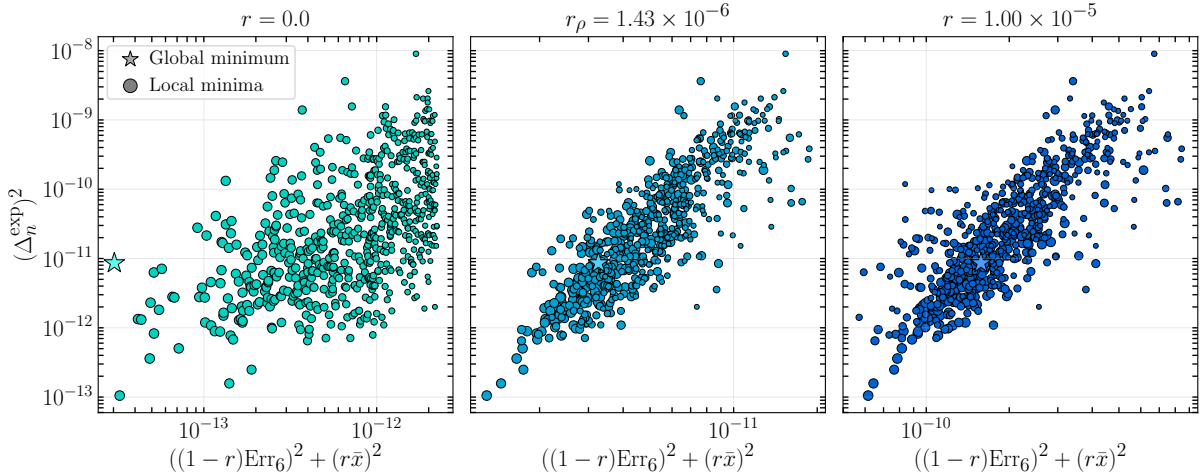}
	  }
	\end{center}
	\caption{Comparison of the Heisenberg XXZ experimental error $\Delta_{n}^{\textrm{exp}}$~\eqref{eq:Frobenius} and the theoretical error~\eqref{eq:theo_err} for found schemes at $q = 14$ cycles and order $n = 6$.
		The correlation of the errors without the added distance from the origin term ($r = 0.0$) is small, which is improved at a certain value of the ratio $r$.
    The sharp cutoff at $r = 0.0$ appears because we exclude the schemes, which fall below a certain efficiency.
		We find maximal correlation at $r_{\rho} = 1.43 \times 10^{-6}$ by computing the Pearson correlation coefficient.
    Above this value the distance from the origin $\bar{x}$ dominates the theoretical error~\eqref{eq:theo_err}, and the correlation decreases.
		An animation of this plot over the values of the ratio $r$ can be found at~\cite{markomalezic_2026_18347430}, and presents how the correlation changes with $r$.}\label{fig:Err_comp}
\end{figure}

\begin{figure}[h!]
	\begin{center}
		\resizebox{0.9\textwidth}{!}{%
		\input{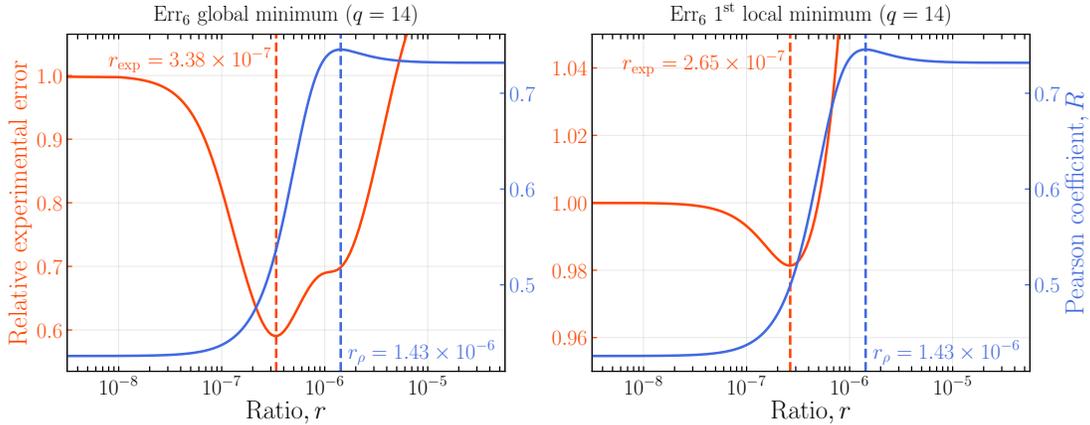}
	  }
	\end{center}
	\caption{Improved relative errors of the global (left) and $1^{\textrm{st}}$ local minima (right) in the experimental $\textrm{Err}_{6}$ function alongside the Pearson correlation coefficient $R$ against the ratio $r$.
          We obtained the Pearson coefficient from Fig.~\ref{fig:Err_comp} and found the maximal correlation at $r_{\rho} = 1.43 \times 10^{-6}$, where we expected to find an improvement.
          After applying the improvement procedure at different ratios $r$, we instead find that the improvement exists, but at a different ratio $r_{\textrm{exp}}$ for both analysed minima.
          The improvement at this ratio is around $40 \%$ for the global minimum, while it is only about $2 \%$ for the $1^{\textrm{st}}$ local minimum.
          We believe closeness to the origin point $\bar{x}_{\textrm{min}}$ impacts the scale of the improvement.}\label{fig:Pearson}
\end{figure}

We performed more numerical experiments on schemes with $q = 14$ cycles and computed the accumulated experimental error $\Delta_{n}^{\textrm{exp}}$~\eqref{eq:Frobenius} for the Heisenberg XXZ model in the global ordering $S^{(3)}$, and propose an improvement to our framework.
Motivated by the origin point constraint, we combine the leading error $\textrm{Err}_{6}$ and the distance from the origin point $\bar{x}$~\eqref{eq:origin},

\begin{equation}
	\textrm{Err}^{2} = \left( (1-r) \textrm{Err}_{n} \right)^{2} + (r \bar{x})^{2},\label{eq:theo_err}
\end{equation}

\noindent
where we introduce a ratio $r \in [0, 1]$ between the two terms.
Comparing the theoretical error~\eqref{eq:theo_err} at three different values of the ratio $r$ to the experimental error we find some interesting results as seen in \Cref{fig:Err_comp}.
In the limit without the additional term $r = 0.0$, where we expected a linear correlation at first, we find a noticeable discrepancy between the two errors.
Here we can also clearly see how the global minimum performs much worse than the first local minimum of the $\textrm{Err}_{6}$ error function.
The sudden cutoff also occurs here due to the exclusion of schemes that fall below our efficiency threshold.
Turning on the extra term with the ratio $r$, we can see the correlation between errors improve, which can be seen better through an animation found at~\cite{markomalezic_2026_18347430}.
Observing this, it turns out that there exists a sweet spot around $r_{\rho} = 1.43 \times 10^{-6}$, where the linear correlation is maximal.
This value can be obtained by minimizing the Pearson correlation coefficient $\rho$,

\begin{equation}
  \rho = \frac{\textrm{cov}(X, Y)}{\sigma_{X} \sigma_{Y}},
\end{equation}

\noindent
where $X$ and $Y$ represent the paired sets of the theoretical~\eqref{eq:theo_err} and experimental error~\eqref{eq:Frobenius}, $\textrm{cov}$ denotes their covariance and $\sigma_{X,Y}$ the respective standard deviations.
We present this in blue in \Cref{fig:Pearson}, where we observe a slight peak in the correlation, which gives the optimal estimate for the ratio $r_{\rho}$ based on the Pearson correlation coefficient.

Returning to the minimization process, and optimizing the new error function~\eqref{eq:theo_err} with some ratio $r$, we now produce schemes with (slightly) perturbed parameters.
We applied this procedure to our global and $1^{\textrm{st}}$ local minimum of the leading-order $\textrm{Err}_{6}$ function at varying ratios $r$.
With the perturbed schemes we again simulate the time evolution and plot the accumulated experimental error relative to the error we accumulate with the original schemes (see Fig.~\ref{fig:Pearson}).
Once again, the practical picture looks different from the theoretically expected one.
Firstly, we do improve the efficiency of our original schemes, and we find a maximal improvement at some ratio $r_{\textrm{exp}}$ for both of our tested schemes.
However, this ratio does not agree with the one calculated from the Pearson correlation coefficient $r_{\rho}$.
It is possible that there still exist some practical effects, which we did not include in the theoretical error.
Secondly, there is a big difference in the improvement of the global and the $1^{\textrm{st}}$ local minima.
At maximal improvement $r_{\textrm{exp}}$ the global minimum performs approximately $40 \%$ better than the original, while the $1^{\textrm{st}}$ local minimum improves only around $2 \%$.
We believe the reason is the proximity of the original schemes to the minimal distance from the origin $\bar{x}_{\textrm{min}}$ (see Fig.~\ref{fig:origin_err}).
We should note that even in the maximal improvement of the global minimum, it does not come close to performing better than the original $1^{\textrm{st}}$ local minimum.
Furthermore, due to the strong model dependence of schemes, we cannot be sure the same sort of improvement can be found in other systems.
It is because of this and the minimal improvement of our recommended $6^{\textrm{th}}$ order scheme, that we find the procedure impractical, though further research might uncover new areas of exploration.

\newpage

%% file: conclusion.tex
% !TEX root = Efficient_Trotterizations

\section{Conclusion}\label{sec:conclusion}

We revisited the construction of higher order Trotter-Suzuki schemes with the intention to improve time evolution in quantum systems.
Following the method by Omelyan \textit{et al.}~\cite{OMELYAN2003272} we built a general framework to find highly efficient decompositions from scratch in \Cref{sec:framework}.
A Trotter scheme of order $n$ is efficient if its leading-order error is small relative to the cost of a single Trotter step.
\Cref{fig:ramp} illustrates how the cost of a Trotter step is directly proportional to the number of cycles $q$.
Error and efficiency are defined in Eqs.~\eqref{eq:order_error} and~\eqref{eq:eff}, respectively.
In \Cref{sec:recursive} we derive a recursive prescription to calculate the leading-order error.
By minimizing the error functions described in \Cref{sec:optim}, we were able to identify the top most efficient Trotterizations at orders $n = 2, 4, 6$ (only the scheme at $n=2$ was known previously~\cite{OMELYAN2003272}).
We found that at each order the maximal number of cycles $q_{\textrm{max}}$ also yields maximal efficiency and thus recommend the use of our scheme at $q = 6$ for order $n = 4$ (see \Cref{tab:recommened6}) and, better yet, the scheme at $q = 14$ for order $n = 6$ (see \Cref{tab:recommened14}).
The recommended schemes were chosen not only because of their high theoretical efficiency but also practical performance in numerical experiments.

First, we experimented on the Heisenberg model in \Cref{sec:novel_schemes:Heisenberg}, by comparing exactly time-evolved states to those approximated by our novel schemes.
We studied two fundamentally different orderings of the Hamiltonian terms, and found similar results to that of the theoretical efficiency (see \Cref{fig:efficiency_Heisenberg}).
In these experiments, we first observed the property that the error accumulation over time appears to be much worse for Trotter schemes with larger distance from the origin (see \Cref{sec:error_accumulation}).
This property led us to recommend the theoretically second most efficient schemes rather than the first.
We then ran more simulations with the two recommended schemes and compared how they perform against historically relevant schemes at fixed final time across the computation cost in \Cref{fig:efficiency_Heisenberg}.
Somewhat surprisingly, we found that our recommended order $n = 6$ scheme performed better than the other order $n \leq 6$ schemes across the relevant computational cost.
This includes the region of lower costs (few large time steps), where lower order schemes were expected to prevail.
It also fared better than the order $n = 8, 10$ schemes by Morales \textit{et al.}~\cite{morales2022greatly} in a significant region of the cost.
Consequently, it is beneficial to use the $6^{\textrm{th}}$ order Trotter scheme from \Cref{tab:recommened14}, almost regardless of the target accuracy (at least for similar models).
To conclude the Heisenberg model results, we looked at how the accuracy changes with the length of the chain $L$, and found that it stabilizes around $L = 6$ (see \Cref{fig:chain}).
This implies that the Trotter scheme that performs best for a given physical system at small system size is expected to consistently remain optimal for large sizes.

We also performed similar numerical experiments for the quantum harmonic oscillator in \Cref{sec:novel_schemes:Harmonic}, which differs from the Heisenberg model in many respects.
Though not surprising, we find quite a model dependence on the performance of Trotter-Suzuki schemes.
Different schemes behave unexpectedly due to different Hamiltonian properties, and the recommended schemes are no longer always optimal.
Nevertheless, they are consistently among the best decompositions.
We believe that the high theoretical efficiency and close proximity to the origin of the recommended schemes ensures universally high accuracy.
It should be noted that if maximal efficiency is desired for a specific model, the optimal scheme parameters can in principle be found for small system size.
This can be then expanded to larger systems as we have shown with the Heisenberg model (see \Cref{fig:chain}).

In \Cref{sec:uniformity} we returned to the distance from the origin property and attempted to improve our framework.
Motivated by empirical observations and the error accumulation arguments from Ref.~\cite{chen2024trottererrortimescaling,chen2024errorinterferencequantumsimulation}, we added a penalty for large deviations from the origin to the theoretical error.
We found that the correlation between this extended theoretical error and the experimental error of the Heisenberg model can be improved at a certain ratio $r$ of the penalty term~\eqref{eq:theo_err} (see \Cref{fig:Err_comp}).
Animations found at~\cite{markomalezic_2026_18347430} visually demonstrate the improvement in the correlation nicely.
Varying the penalty $r$ for the global and $1^{\textrm{st}}$ local minimum, we were able to improve the practical performance of these schemes as can be seen in \Cref{fig:Pearson}.
However, the optimal values for $r$ did not match and we find minimal improvement in the recommended scheme.
Therefore, we believe the procedure to be impractical in general.
More theoretical insight in this area is called for and we hope to explore it further.

With the framework for higher order scheme construction established, we intend to continue the exploration of efficient Trotter-Suzuki schemes at order $n = 8$.
\Cref{tab:orders} implies that orders higher than $8^{\textrm{th}}$ are likely to not have any free parameters prone to optimisation and could prove inefficient, but we also intend to look into this.
Since we only focused on schemes with real parameters due to their unitarity, there exists a whole area of research into non-unitary schemes, which is made accessible through our framework.
The scheme parameters in this area are complex valued, and thus the minimization problem becomes twice as large.
However, there is evidence to suggest that such schemes perform much better, although they are only restricted to use in non-unitary simulations like those done on classical computers.

Another avenue that we wish to explore is the performance of higher order schemes on quantum hardware.
Based on the results we got from the Heisenberg model, we expect that fewer local gates are needed to accumulate the same Trotter error at a fixed final time, which would decrease the overall hardware noise and improve quantum simulations.
Finally, Trotter-Suzuki schemes can be used for symplectic integration of classical equations of motion, though their efficiency might differ significantly at higher orders because some contributions to the error vanish exactly.
Studying them explicitly might prove useful for simulations of molecular dynamics.
There is clearly still much to be discovered about Trotterizations at higher orders, and we intend to pursue these leads in order to improve simulations of time evolution.

%% file: framework_details.tex
% !TEX root = Efficient_Trotterizations

\section{Framework technical details}\label{sec:framework_details}

In this appendix we expand on some of the more technical details of our framework, specifically those pertaining to the minimization part.
The optimization algorithm we chose in our simulations was the one developed by Levenberg and Marquardt~\cite{Levenberg:1944,Marquardt:1963}.
In order to obtain scheme parameters $a_{i}$ and $b_{i}$, we must minimize the following $\chi^{2}$ function,

\begin{equation}
  \chi^{2} (a_{i}, b_{i}) = \sum_{k=0}^{n} w_{k} \textrm{Err}_{k}^{2} (a_{i}, b_{i}) = \mathbf{y} (\mathbf{p})^{\dagger} \mathbf{W} \mathbf{y} (\mathbf{p}),\label{eq:min_chi2}
\end{equation}

\noindent
which can be rewritten in terms of algebraic notation by defining the coefficient vector $\mathbf{y} = (\nu, \sigma, \alpha, \beta, \gamma_{1}, \ldots)$ and the weight matrix $\mathbf{W} = \mathrm{diag}(w_{0}, w_{0}, w_{2}, w_{2}, w_{4}, \ldots)$.
We also group the scheme parameters in a vector $\mathbf{p} = (a_{1}, a_{2}, \ldots, a_{\nicefrac{q}{2}+1}, b_{1}, b_{2}, \ldots, b_{\nicefrac{q}{2}})$, where we assumed a scheme with an even number of cycles $q$.
We note here that the scheme parameters in $\mathbf{p}$ are considered complex-valued, which also makes the scheme coefficients inside $\mathbf{y}$ complex.
Besides the $\chi^{2}$ function, the algorithm needs to know its gradient,

\begin{equation}
  \frac{\partial}{\partial \mathbf{p}} \chi^{2} = 2 \mathbf{y} (\mathbf{p})^{\dagger} \mathbf{W} \mathbf{J}, \quad \mathbf{J} = \frac{\partial \mathbf{y} (\mathbf{p})}{ \partial \mathbf{p}},\label{eq:chi2_grad}
\end{equation}

\noindent
where we defined the Jacobian matrix~$\mathbf{J}$ of the scheme coefficients~$\mathbf{y}$ over their parameters~$\mathbf{p}$.

With this we can now propose a step $\mathbf{h}_{\textrm{LM}}$ in the manifold of $\chi^{2} (\mathbf{p})$ over the parameter space $\mathbf{p}$, according to ideas by Levenberg and Marquard,

\begin{equation}
  \left[ \mathbf{J}^{\dagger} \mathbf{W} \mathbf{J} + \lambda\, \textrm{diag}(\mathbf{J}^{\dagger} \mathbf{W} \mathbf{J}) \right] \mathbf{h}_{\textrm{LM}} = - \mathbf{J}^{\dagger} \mathbf{W} \mathbf{y} (\mathbf{p}),\label{eq:LM_step}
\end{equation}

\noindent
where we introduced the damping parameter $\lambda$.
This allows the combination of the Gauss-Newton method at small ($\lambda < 1$) values of $\lambda$, and the gradient descent at large values ($\lambda > 1$).
The matrix that appears on the left side of Eq.~\eqref{eq:LM_step} roughly approximates the Hessian $\mathbf{H} \approx \mathbf{J}^{\dagger} \mathbf{W} \mathbf{J}$, which has an important role.
Further details on the acceptance condition of the step $\mathbf{h}_{\textrm{LM}}$ and the convergence conditions can be found in Ref.~\cite{gavin2024levenberg}.

\subsection*{Optimizer implementation}

Our minimization function $\chi^{2}$ is composed of scheme coefficients $\mathbf{y} (\mathbf{p}) = (\nu, \sigma, \alpha, \beta, \gamma_{1}, \ldots)$, and for a given number of parameters $\mathbf{p}$ it is possible to fully set a coefficient to zero.
In other words, this means applying the constraints, which satisfy a desired order $n$.
The $0^{\textrm{th}}$ order constraints are as follows, $\nu = 2 \sum_{i} a_{i} = 1$ and $\sigma = 2 \sum_{i} b_{i} = 1$.
These two constraints can be exactly enforced by fixing one of the parameters of $a_{i}$ and $b_{i}$, and we choose to fix the final ones:

\begin{equation}
  a_{\nicefrac{q}{2}+1} = \frac{1}{2} - \sum_{i}^{\nicefrac{q}{2}} a_{i}, \quad b_{\nicefrac{q}{2}} = \frac{1}{2} - \sum_{i}^{\nicefrac{q}{2}-1} b_{i}.\label{eq:fixed_params}
\end{equation}

\noindent
This way we effectively reduce the minimization problem in two ways.
There are 2 fewer parameters to minimize, and the coefficient vector can be redefined to $\mathbf{y} = (\alpha, \beta, \gamma_{1}, \ldots)$, where we removed the coefficients $\nu$ and $\sigma$.
However, there is a caveat when it comes to computing the Jacobian $\mathbf{J}$, which is also reduced, but one has to take care when computing its derivatives.
Because the final parameter $a_{\nicefrac{q}{2}+1} (a_{i})$ depends on the rest of the scheme parameters $a_{i}$ one needs to use the chain rule on the $\chi^{2} (\mathbf{p})$ in the following way,

\begin{equation}
  \frac{\partial}{\partial a_{i}} \chi^{2} \left( \mathbf{p} (a_{i}, a_{\nicefrac{q}{2}+1} (a_{i})) \right) = \frac{\partial \chi^{2}}{\partial a_{i}} + \frac{\partial \chi^{2}}{\partial a_{\nicefrac{q}{2}+1}} \frac{\partial a_{\nicefrac{q}{2}+1}}{\partial a_{i}}, \quad \frac{\partial a_{\nicefrac{q}{2}+1}}{\partial a_{i}} = -1.\label{eq:final_param_a}
\end{equation}

\noindent
and similarly for the final parameter $b_{\nicefrac{q}{2}} (b_{i})$,

\begin{equation}
  \frac{\partial}{\partial b_{i}} \chi^{2} \left( \mathbf{p} (b_{i}, b_{\nicefrac{q}{2}} (b_{i})) \right) = \frac{\partial \chi^{2}}{\partial b_{i}} + \frac{\partial \chi^{2}}{\partial b_{\nicefrac{q}{2}}} \frac{\partial b_{\nicefrac{q}{2}}}{\partial b_{i}}, \quad \frac{\partial b_{\nicefrac{q}{2}}}{\partial b_{i}} = -1.\label{eq:final_param_b}
\end{equation}

\noindent
Thus, we do not avoid computing all the derivatives of $\chi^{2}$ over the parameters in $\mathbf{p}$.
However, exactly fixing the $0^{\textrm{th}}$ order constraint allows us to better impose higher orders, and a reduction in the space of the parameters and coefficients improves the cost of our simulations.

Imposing higher orders includes further technical details, which are non-trivial.
Firstly, appropriate values for the weights need to be found empirically, and the specific values used in our simulations are collected in \Cref{tab:weights}.
Since the weights are relative to each other, we can fix one of them, e.g.\ the weight at order $n = 2$ to $w_{2} = 1.0$.
The rest of them are then experimented with such that the constraints are satisfied to numerical precision, while keeping the leading-order error small.
However, simply setting the weights per order and minimizing in regions with free parameters (see Tab.~\ref{tab:orders}) will, depending on the choice of the weights, only approximately impose the constraints.
Therefore, after minimizing once, a second optimization step is needed, which fully imposes the constraint by ignoring the leading-order errors.
The idea is to approach the region of the leading-order error minimum in the first step, and converge to the valid solution at that order, by imposing the constraint.
In order to do this effectively, we `freeze' the Hessian $\mathbf{H}$ after the first step, which helps to stay in the vicinity of the leading-order error minimum.
Without the freezing, steps in the unconstrained directions might be overly large.
In the parameter regions without free parameters (see Tab.~\ref{tab:orders}), no such procedure is needed, because the imposing of constraints is the only step needed.
As a final note, we remark that imposing the constraints $\textrm{Err}_{k} = 0$ is only possible to a desired numerical precision.
In our simulations the precision was 18 significant digits, although this can be improved without much effort.

%\begin{table}[h!]
%	\centering
%	\caption{The table collects the specific weights used for minimization in different regions of cycles $q$.
%           The first weight $w_{2}$ can be fixed to 1.0, because the important value is relative to the other weights $w_{k}$.
%           The other weights are experimented with in order to find a good region, where numerical precision for the constraints is achieved, and the leading-order error is minimized well.
%           The parentheses denote ranges of integer values.}
%	\begin{tabular}{ccccc}
%		\toprule
%		\multicolumn{1}{c}{Order $n$}  & \multicolumn{1}{c|}{2}      & \multicolumn{1}{c|}{4}      &                   \multicolumn{2}{c|}{6}                           \\
%		\multicolumn{1}{c}{Cycles $q$} & \multicolumn{1}{c|}{[1, 2]} & \multicolumn{1}{c|}{[3, 6]} & \multicolumn{1}{c|}{[7, 8]} & \multicolumn{1}{c|}{[9, 14]}     \\
%		\midrule
%    \multicolumn{1}{c}{$w_{2}$}    & \multicolumn{1}{S|}{1.0}    & \multicolumn{1}{S|}{1.0}    & \multicolumn{1}{S|}{1.0}        & \multicolumn{1}{S|}{1.0}         \\
%    \multicolumn{1}{c}{$w_{4}$}    & \multicolumn{1}{S|}{\color{gray}0.0}       & \multicolumn{1}{S|}{0.1}    & \multicolumn{1}{S|}{100.0}      & \multicolumn{1}{S|}{500.0}       \\
%    \multicolumn{1}{c}{$w_{6}$}    & \multicolumn{1}{S|}{\color{gray}0.0}       & \multicolumn{1}{S|}{\color{gray}0.0}       & \multicolumn{1}{S|}{\color{gray}0.0}           & \multicolumn{1}{S|}{1.0} \\
%		\bottomrule
%	\end{tabular}\label{tab:weights}
%\end{table}

\begin{table}[h!]
	\centering
	\caption{The table collects the specific weights used for minimization in different regions of cycles $q$.
           The first weight $w_{2}$ can be fixed to 1.0, because the important value is relative to the other weights $w_{k}$.
           The other weights are experimented with in order to find a good region, where numerical precision for the constraints is achieved, and the leading-order error is minimized well.
           The parentheses denote ranges of integer values.}
	\begin{tabular}{ccccc}
		\toprule
		\multicolumn{1}{c}{Order $n$}  & \multicolumn{1}{c|}{2}      & \multicolumn{1}{c|}{4}      &                   \multicolumn{2}{c|}{6}                           \\
		\multicolumn{1}{c}{Cycles $q$} & \multicolumn{1}{c|}{[1, 2]} & \multicolumn{1}{c|}{[3, 6]} & \multicolumn{1}{c|}{[7, 8]} & \multicolumn{1}{c|}{[9, 14]}     \\
		\midrule
    \multicolumn{1}{c}{$w_{2}$}    & \multicolumn{1}{r|}{1.0}    & \multicolumn{1}{r|}{1.0}    & \multicolumn{1}{r|}{1.0}        & \multicolumn{1}{r|}{1.0}         \\
    \multicolumn{1}{c}{$w_{4}$}    & \multicolumn{1}{r|}{\color{gray}0.0}       & \multicolumn{1}{r|}{0.1}    & \multicolumn{1}{r|}{100.0}      & \multicolumn{1}{r|}{500.0}       \\
    \multicolumn{1}{c}{$w_{6}$}    & \multicolumn{1}{r|}{\color{gray}0.0}       & \multicolumn{1}{r|}{\color{gray}0.0}       & \multicolumn{1}{r|}{\color{gray}0.0}           & \multicolumn{1}{r|}{1.0} \\
		\bottomrule
	\end{tabular}\label{tab:weights}
\end{table}

%% file: error_accumulation.tex
% !TEX root = Efficient_Trotterizations

\section{Minimal error accumulation in 2nd order schemes}\label{sec:error_accumulation}

Following Ref.~\cite{chen2024trottererrortimescaling}, the Trotter error~\eqref{eq:even_symm_scheme} can be split into operators `parallel` and `orthogonal` to the full Hamiltonian $H$.
Contributions in the parallel subspace $H_\parallel$ add up linearly over time while those in the orthogonal subspace $H_\perp$ change direction permanently and do not accumulate.
With local $\order{h^{n+1}}$ errors this leads to an overall error scaling with $\order{h^n t}+ \order{h^n}$.
Clearly, for long time evolutions it is advantageous to reduce or even eliminate the leading $\order{h^n t}$ error stemming from $H_\parallel$.

For second order schemes it is possible to split the error operators according to their subspaces exactly.
We write the $q=2$ cycle (i.e.\ the largest 2nd order) Trotterization for two operators as
\begin{align}
	e^{\lambda A h} e^{B h/2} e^{(1-2\lambda) A h} e^{B h/2} e^{\lambda A h} &= e^{(A+B) h + \alpha [A,[A,B]] h^3 + \beta [B,[A,B]] h^3 + \order{h^5}}
\end{align}
with the single free parameter $\lambda$ and the coefficients~\cite{OMELYAN2003272}
\begin{align}
	\alpha = \frac{1}{12} - \frac{\lambda}{2} + \frac{\lambda^2}{2}\,,\ \beta = \frac{1}{24} - \frac{\lambda}{4}\,.
\end{align}
The leading-order error can be rewritten using $A+B=H$
\begin{align}
	O_3 &= \left(\alpha-\beta\right) [A,[A,B]] + \beta [H,[H,B]]\,,
\end{align}
where the second term $[H,[H,B]]\in H_\perp$ does not contribute to the error accumulation.

In order to eliminate leading-order error accumulation entirely, thus elevating the total error to $\order{h^4 t}+ \order{h^2}$, we need to satisfy
\begin{align}
	\alpha-\beta &= 0\\
	\Leftrightarrow \lambda &= \frac14 \left(1 \pm \frac{i}{\sqrt{3}}\right)\,.
\end{align}
Unfortunately, there are no real solutions, but this scheme with complex-valued coefficients might prove highly efficient in practice.

The unique real minimum of the parallel contribution is reached for $\lambda=\frac14$, that is for uniform coefficients.
In fact, $\lambda=\frac14$ makes this decomposition equivalent to two equal Leapfrog steps.
Projecting this result to higher order Trotter schemes, we expect the error accumulation to be generally preferable for maximally uniform coefficients.